\documentclass{jfm}
\usepackage{graphicx}
\usepackage{caption}
\usepackage{subcaption}
\usepackage{epstopdf, epsfig}
\usepackage{amsmath}
\usepackage{cleveref}
\usepackage{float}
\usepackage{tikz}
\usetikzlibrary{positioning}
\usetikzlibrary{patterns.meta}
\usetikzlibrary{decorations,decorations.markings,decorations.text}
\usetikzlibrary{decorations.pathmorphing,arrows,intersections}
\usetikzlibrary{shadings}
\usepgflibrary{shadings}

\usepackage{tikz-dimline}
\newcommand*{\addheight}[2][.5ex]{%
  \raisebox{0pt}[\dimexpr\height+(#1)\relax]{#2}%
}

\title{Migration of two Interacting Micro-Confined Deformable Drops Under
an Imposed Temperature Gradient}

\author{Sayak Ray
  \corresp{\email{sayakray@kgpian.iitkgp.ac.in}},
  Sudipta Ray
 \and Prof. Suman Chakraborty}

\affiliation{\aff{1}Department of Mechanical Engineering, Indian Institute of Technology, Kharagpur,
Kharagpur, India}
\pgfplotsset{compat=1.18} 
\begin{document}

\maketitle

\begin{abstract}
Thermocapillary motion is widespread in both natural occurrences and engineering applications. A tiny drop of one liquid, suspended within another, may be set into motion aligned with an imposed thermal gradient, as influenced by thermocapillary action stemming from the gradients in interfacial tension due to the local variations in temperature. In real-world situations, however, such drops do not remain in isolation, as they interact with their neighbouring entities including other drops in the proximity as well as a nearby solid boundary, setting up a complex interplay between the confinement-mediated interactions and three-dimensional nature of the droplet dynamics. In this study, we present numerical solutions for the migration dynamics of a tightly-confined drop-couple, incorporating deformable interfaces, film flow, and Marangoni effects in the presence of dynamically evolving thermocapillary stresses induced by an imposed uniform temperature gradient. Unlike prior investigations, our work highlights the influence of the confinement towards orchestrating non-trivial features of drop migration, as dictated by an intricate coupling of the thermal and flow fields amidst the interferences of the domain boundaries. The study reveals that hydrodynamic interactions resulting from a juxtaposition of these influences deform the drops in a unique manner as compared to the characteristics evidenced from previously reported studies, causing a distortion of the local thermal fields around them. This, in turn, leads to changes in the local thermocapillary stress, affecting the local shear gradient in a manner that alters the local flow field in accordance to ensure the interfacial stress balance. The consequent alteration in the drop velocities is shown to govern their migration in a distinctive manner, presenting unique signatures as compared to more restrictive scenarios studied previously. These findings hold significance in designing thermocapillary-driven micro-confined systems for controlling drop trajectories under an imposed thermal field, bearing far-reaching implications in a plethora of overarching applications ranging from droplet microfluidics to space technology.
\end{abstract}

% \begin{keywords}
% Authors should not enter keywords on the manuscript, as these must be chosen by the author during the online submission process and will then be added during the typesetting process (see http://journals.cambridge.org/data/\linebreak[3]relatedlink/jfm-\linebreak[3]keywords.pdf for the full list)
% \end{keywords}

\section{Introduction}
Thermocapillary motion is pervasive in both natural phenomena and engineering applications. A small drop of one fluid, suspended in another medium having a prevailing temperature gradient, may exhibit movement in the direction of the imposed gradient, as dictated by a resulting thermocapillary action. [REFS]. This phenomenon is typically orchestrated by the fact that the presence of a local temperature gradient generates a corresponding gradient of the interfacial tension along the surface of the drop. This differential tension acts as a force, pulling the surrounding fluid and propelling the drop towards areas where its interfacial tension would typically be lower, often in the direction of the hotter regions. Such phenomena, which have been intriguing physicists over the years, have become progressively more important over the past decades from their application-oriented perspectives due to unprecedented advancements in miniaturization and space technology where several utilities are to function in near weightless conditions. For example, the removal of unwanted liquid drops in a continuous phase by thermocapillary forces may greatly facilitate the processing of materials in outer space, minimizing various defects that are otherwise inevitable due to gravity-induced fluid phase segregation  \cite[]{uhlmann1981,carruthers1983,ostrach1982}. It is not far beyond imagination that the cooling system of space habitats may be achievable using thermocapillary migration. With the advent of microfluidics and miniaturization, thermocapillary phenomena have been attracting attention in several on-earth applications as well, such as micro heat pipes for thermal management of electronic equipment [REFS], where gravity-induced effects render to be inconsequential due to their large surface area by volume ratios. In several such scenarios, uncontrolled accumulation of drops via thermocapillary motion may be rather undesirable, as they may deteriorate the heat exchange efficacy between the hot and the cool interfaces. Imposing a delicate control over the thermocapillary migration of drops, therefore, appears to be imperative, irrespective of whether their motion needs to be accelerated or retarded.
A vast body of reported research on thermocapillary motion concerns the migration of single drops or bubbles in isolation \cite[]{young1959,hetsroni1970,Balasubramaniam1987,haj1997thermocapillary,chan1979,haj1990,zhang2001,wu2012}. For an accounting of the early studies in this field, one may refer to the review papers by \citet{subramanian2002} and \citet{wozniak1988}. While early investigations on this topic considered drops in unbounded flows, subsequent endeavors probed more closely the effects of the confining walls \cite[]{brady2011} on the drop dynamics. A noteworthy finding regarding the impact of boundary effects on thermocapillary motion was that a drop with significant thermal conductivity can undergo faster migration near a free fluid surface compared to when it is isolated. Nevertheless, in real-world applications, managing numerous bubbles or drops is often essential, and their collective behavior may deviate significantly from the intuitive expectations based on the individual particle outcomes, so that understanding the dynamics of interacting drops renders to be practically more imperative. 
In several practical scenarios, a drop interacts simultaneously with the domain boundaries and other neighbouring drops \cite[]{keh1992}. These interactions may introduce strong local variations in the temperature gradients on the interfaces of the drops, leading to localized changes in the surface tension. The consequent alterations in the interfacial surface tension gradient-driven fluid motion near the interfaces may perturb the drop’s shape evolution and motion simultaneously, as the interfaces are drawn in the direction of increasing interfacial tension. The consequent interaction of the drops, in lieu, may perpetually modify the local interfacial interactions (stress jump conditions) in a manner that may cause significant drop deformations even in the case of negligible convective transport. Further to this end, when attracted in sufficiently close vicinity by virtue of local gradients in interfacial tension, the interacting drops may coalesce as well, as observed in different natural and industrial processes, including liquid–liquid phase separation, polymer casting, and the treatment of liquid phase-miscibility-gap materials. For situations in which such coalescence renders undesirable such as the ones considered in this work, a careful a-priori rationalization of the drop interaction dynamics renders critical, in line with the intended drop migration features consistent with the particular application on focus.
In the literature, initial studies on the interaction of drops in the course of their thermocapillary migration were performed under the assumption of negligible deformation (capillary number tends to zero in the limit as applicable for quiescent flows of large surface tension or low viscosity fluids); for example, see the articles of \citet{meyyappan1984,meyyappan1983} and \citet{acrivos1990}. The motion of two liquid drops oriented arbitrarily with respect to a temperature gradient was examined analytically by \citet{anderson1985} in the low Reynolds and Marangoni number limit.  By using the two-drop solution, he also showed that the mean velocity of a drop suspension is lower than for a single drop. \citet{keh1990} examined the axisymmetric thermocapillary motion of two spherical drops progressing along their line of centres within a creeping flow. Their findings demonstrated that two identical liquid drops exhibit a faster migration compared to a single drop of the same size. Conversely, in the case of two gas bubbles with equal radii, no interaction was observed for all separation distances, aligning with the predictions made by \citet{meyyappan1984}. Later \citet{keh1992} explored the axisymmetric migration of a series of spherical drops and gas bubbles moving along their line of centres. In the case of multiple gas bubbles, it was demonstrated that the migration velocity of each bubble remained unaffected by the presence of the other bubbles, if the bubbles are of the same size. \citet{wei1993} investigated theoretically the quasi-static thermocapillary migration of a chain of two and three spherical bubbles for zero Marangoni and Reynolds numbers. \citet{keh1992} considered the migration of drops oriented arbitrarily with respect to the temperature gradient in the limit of zero Marangoni and Reynolds numbers. Unlike drops moving along their line of centers \cite[]{keh1990} drops moving with their line of centers orthogonal to the temperature gradient were shown to migrate slower than a single drop. \citet{loewenberg1993} studied the axisymmetric, thermocapillary-driven motion of a pair of non-conducting, spherical drops in near-contact for small Reynolds and Marangoni numbers. Their study involved computing the pairwise motion and associated contact forces by examining touching drops in point-contact. In this scenario, the relative motion between nearly touching drops initiated from the contact force, which was counteracted by a lubrication resistance. The conclusion drawn was that, for nearly equisized drops, the ratio of relative velocity between two drops in near contact to that for widely separated drops is consistent for both thermocapillary-driven and gravity-driven motion.
The interaction of two deformable drops in the axisymmetric coordinates was studied by \citet{zhou1996}. Numerical simulations of an axisymmetric buoyancy-driven interaction of a leading drop and a smaller trailing drop were reported by \citet{zinchenko1999}. This study revealed that the trailing drop experiences significant elongation due to the hydrodynamic influence exerted by the leading drop. Subsequently, depending on the governing parameters, the drops may either separate and revert to a spherical shape, the trailing drop may be captured by the leading one, or one of the drops may undergo breakup. In the context of thermocapillary-induced motion, the impact of deformability was primarily investigated using a perturbation technique, assuming small deformations. \citet{rother1999} applied lubrication approximation to study the effect of a slight deformability of the interfaces on the thermocapillary driven migration of two drops at close proximity.
The investigations of interactions of drops discussed above have all been limited to zero Reynolds and Marangoni numbers. \citet{nas2003} conducted a computational investigation into the thermocapillary migration of two fully deformable bubbles and drops, considering non-zero values of the Reynolds and Marangoni numbers. The results indicated that the bubbles and lightweight drops got aligned perpendicularly to the temperature gradient and were uniformly spaced in the horizontal direction. A space experiment evidenced that a small leading drop could retard the movement of the big trailing drop in the process \cite[]{balasubramaniam1996}. \citet{yin2011} studied the thermocapillary interaction of two arbitrarily placed drops, considering them to be of the same size and having the same physical parameters (kinematic viscosity, thermal diffusivity, density, and specific heat).
One critical feature that was shown to demarcate the characteristics of two interacting drops as compared to the corresponding single-drop dynamics is the distinction between their impending coalescing regime and in-tact motion. Of great interest is the non-coalescing behaviour of the interacting drops, which had its early foundation on the seminal studies of Lord Rayleigh on the behaviour of water jets that bounce over one another \cite[]{rayleigh1899}, with its resurgence about a century later in the form of surface vibration-facilitated non-coalescence \cite[]{walker1978} that fundamentally aimed to inhibit the impending drainage of liquid between the two interacting drops in close proximity  \cite[]{marrucci1969,anilkumar1991}. Systems exhibiting this apparently unusual non-coalescence thence continued to attract attention, particularly for their implications in materials science, meteorology and microgravity experiments \cite[]{fredriksson1984}. Under thermocapillary effects, the enhanced interfacial shear due to Marangoni effects, opposing the draining of the film between two interfaces, may resist the interfacial tension that facilitates coalescence, resulting in a dynamic enhancement of the resistance to drainage. It is, therefore of no surprise that thermocapillary effects are among the possible means that may be harnessed to result in controlled movement of multiple interacting drops without having them intermingled \cite[]{dell1996}. However, addressing such problems from a theoretical perspective remains challenging, as attributable to the non-negligible drop deformation, three dimensionalities of the transport stemming from possible confinement-induced interactions, and a dynamic alteration in the thermocapillary stress field because of spontaneously varying temperatures around the interacting drops despite applying an imposed uniform thermal gradient. Most imperatively, these key effects do not act in isolation but have a non-trivial interplay because of a two-way coupling between the heat transfer and fluid flow as mediated by an interfacial stress balance that delves on a dynamically evolving thermal field around the interacting drops. Whereas the analytical techniques put forward to addressing the dynamics of a droplet couple promised to be richly insightful to an extent, they appeared to be clearly inadequate in deciphering the resulting complex coupling, establishing the need of more exhaustive computational frameworks.
For moderate and large drop deformations, the boundary-integral method, as pioneered by \citet{rallison1978} and described in detail by \citet{pozrikidis1992}, emerged to be of utilitarian importance for analyzing the hydrodynamic problem in the Stokes flow limit. \citet{zhou1996} conducted a study on the asymmetric motion of two deformable drops subjected to a temperature gradient along the line of their centres. Disregarding heat convection and inertial effects, they computed the temperature and velocity fields for significant drop deformations using boundary-integral techniques for the Laplace and the Stokes equations, respectively. They presented detailed numerical results on drop motion, deformation, and the temporal evolution of the gap width between the drops, considering equal viscosities of the drops and surrounding fluid. The study highlighted the influence of the capillary number, drop size ratio, and drop-to-medium conductivity ratio on drop motion and deformation. Their results indicated that hydrodynamic interactions between the drops exerted a more pronounced effect on the smaller of the two drops, impacting both drop motion and deformation. Deformation was shown to significantly affect the rate of thin film drainage between the drops, while its impact on the velocities of the drop centers was relatively marginal. To the limit of their calculations, they were able to confirm the predictions of \citet{loewenberg1993}. \citet{berejnov2001} analyzed the problem with the trailing drop smaller than or equal to the leading drop. \citet{zhou1996} used a boundary-integral technique to study the thermocapillary interaction of a deformable viscous drop with a larger trailing drop making no a priori assumptions regarding the magnitude of deformations, inferring that the influence of deformability became significant only when the drops came close to each other. However, they did not consider the variations in the surface tension due to continuous changes in the positions of the drops. \citet{berejnov2001} investigated the influence of the deformations on the relative motion of the drops in the case of moderate capillary numbers and equal viscosity and thermal properties of the dispersed and continuous phases.  \citet{rother2002} generalized the above axisymmetric analyses to three dimensions and arbitrary viscosity ratio by adapting the boundary-integral code of \citet{zinchenko1999} to handle the tangential Marangoni stresses. \citet{lavrenteva2001} probed the scenarios of high Peclet number for analyzing the thermocapillary interaction among drops. However, the confinement effects on two-drop thermocapillary interaction amidst a dynamically evolving Marangoni stress acting on them remained to be addressed thus far.
Here we arrive at three-dimensional numerical solutions of the Stokes equation, with deformable interfaces, film flow, and the Marangoni effects in the presence of dynamically evolving thermocapillary stresses on the application of a uniform temperature gradient on a micro-confined drop pair. In contrast to previous investigations, our work puts forward the effects of confinement amidst the coupled thermal and flow fields, in a boundary element framework. The hydrodynamic interactions due to the confining boundaries are shown to deform the drops from their respective equilibrium shapes, which results in a distortion of the local thermal field around their neighborhoods. This, in turn, alters the local thermocapillary stress, and consequently, the local shear gradient to ensure the interfacial stress balance. The resulting alteration in the flow field is shown to dictate the migration of the drops in an intriguing manner having distinctive signatures as compared to other more restrictive scenarios studied previously. These results are likely to be imperative in designing thermocapillary-driven micro-confined systems for controlled drop trajectories under an imposed thermal field.
 
\section{PROBLEM FORMULATION}
\subsection{Problem description}
\begin{figure}
\begin{center}
            \begin{tikzpicture}[xscale = 1.5, yscale =2, plotmark/.style = {%
           draw, fill=red, circle, inner sep=0pt, minimum size=4pt}]
        \draw[black,very thick] (0,0) rectangle (5,2);
        \shade[shading=color wheel white center] (2,1.53) circle (0.2);
        \shade[shading=color wheel white center] (1,1.3) circle (0.2);
        \draw[name path=curve] (0,0) .. controls (0.5,1) .. (0,2);
        \foreach \y in {0.2,0.4,...,1.9}{
          \path[name path=horizontal] (0,\y) -- + (3,0);
          \draw[-stealth,name intersections={of=curve and horizontal}] (0,\y) -- (intersection-1);}
        \draw[->,thick] (0.0,1) -- (0.65,1) ;
        \draw[->,thick] (0.0,1) -- (0.0,2.2)node[above] {$\overline{z}$};
        \draw[->] (2,1.53) -- (2.3,1.53);
        \draw[->] (2,1.53) -- (2,1.8);
        \draw[->] (1,1.3) -- (1.3,1.3);
        \draw[->] (1,1.3) -- (1,1.6);
        \draw (0.66, 0.85)  node {$\overline{x}$};
        \draw (2.44, 1.5) node {\small $\overline{u}_{x2}$};
        \draw (2.13,1.84) node {\small $\overline{u}_{z2}$};
        \draw (1.4,1.2)  node {\small $\overline{u}_{x1}$};
        \draw (1.14,1.56) node {\small $\overline{u}_{z1}$};
        \coordinate (A) at (-0.3,0);
        \coordinate (B) at (3,0);
        \coordinate (C) at (3,2);
        \coordinate (D) at (-0.3,2);
        \coordinate (E) at (4.2,0);
        \coordinate (F) at (4.2,2);
        \draw (A)--(D);
        % \draw (E)--(F);
        \coordinate (A1) at (2,1.3);
        \coordinate (B1) at (1.65,1.3);
        \coordinate (D1) at (1.65,1);
        \coordinate (A2) at (2,1);
        \coordinate (B2) at (2,1.53);
        \coordinate (ce1) at (1,1.3);
        \coordinate (ce2) at (2,1.53);
        % \draw (A1)--(B1);
         \dimline[label style ={above=0.00004ex,font=\scriptsize}]{(A)}{(D)}{$H$};
         % \dimline[label style = {above=0.00004ex,font=\scriptsize}]{(E)}{(F)}{$T = T_{2}$};
         % doesn't need all these dimension lines, but maybe add one to indicate the transverse offset
         %\dimline[label style ={above=0.00004ex,font=\scriptsize}]{(A1)}{(B1)}{d};
         %\dimline[label style ={above=0.00004ex,font=\scriptsize}]{(B1)}{(D1)}{e};
         \draw[dashed] (1,1.3) -- (2,1.3);
         \draw[->](D1)--(B1) node [midway,anchor=west]{\small $e$};
         \draw[->](A2)--(B2) node [midway,anchor=west]{\small $p$};
         \draw[->](ce1)--(ce2) node [midway,anchor=south]{\small $d$};
        \draw[|->|, rotate around={38:(1,1.3)}] (1,1.3) -- (1.2,1.3)  node [right]{\small $a$};
        \draw[dashed] (-0.1,1) -- (5.1,1);
        \foreach \x in {0.0,0.05,...,5.05}{
          \draw[rotate around={-45:(\x,2)}] (\x,2) -- (\x,2.08);}    
        \foreach \x in {0.0,0.05,...,5.05}{
          \draw[rotate around={-135:(\x,0)}] (\x,0) -- (\x,0.08);}   
        \end{tikzpicture}   .
        \end{center}
\caption{Schematic diagram of the confined channel with two droplets in an aligned arrangement}
\label{main_pic}
\end{figure}
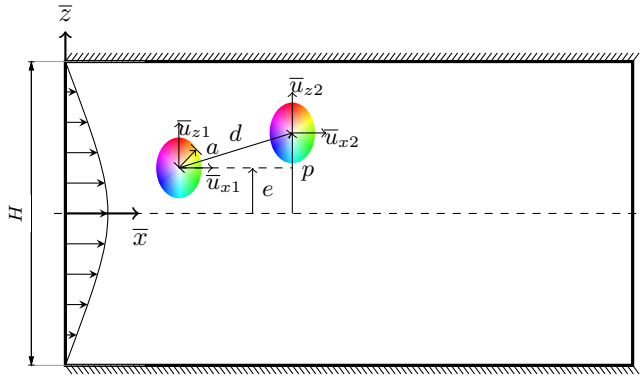

We consider two Newtonian droplets of density $\overline{\rho_i}$, thermal conductivity $\overline{k_i}$, viscosity $\overline{\mu_i}$ in a fluid of density $\overline{\rho_e}$, thermal conductivity $\overline{k_e}$, viscosity $\overline{\mu_e}$. The initial shape of the droplet is given as spherical radius $\overline{a}$. The cartesian coordinate system ($\overline{x},\overline{y},\overline{z}$) is fixed to the centreline of the channel. The initial position of the droplets is at the position ($\overline{x_{c1}}$,$\overline{y_{c1}}$,$\overline{z_{c1}}$) and ($\overline{x_{c2}}$,$\overline{y_{c2}}$,$\overline{z_{c2}}$). The imposed free velocity and temperature are given as $\overline{U_\infty}$ and $\overline{T_\infty}$. The velocity of the droplet centroid is given as ($\overline{U_{x1}}$,$\overline{U_{y1}}$,$\overline{U_{z1}}$) and ($\overline{U_{x2}}$,$\overline{U_{y2}}$,$\overline{U_{z2}}$). 
The schematic of the system is represented in \cref{main_pic}. 
It is assumed a temperature gradient of G is applied to the continuous phase. The surface tension $\overline{\sigma}$ decreases linearly with droplet surface temperature ($\overline{T_s}$). The surface tension at the reference temperature of $\overline{T_\infty}$ is given as $\overline{\sigma_{0}}$

\subsection{Normalization}
Normalization of the space is essential to gain an understanding of the functional parameters affecting the problem. The length scale is given as the channel height ($\overline{H}$). The velocity scale is taken to be the centreline velocity $\overline{V_c}$ while the temperature scale is taken to be $\overline{G}\overline{H}$ as the scaled temperature difference $\overline{\Delta T}$. The corresponding non-dimensional parameters formed are given below 
\begin{enumerate}
    \item Capillary number given as $Ca = \frac{\mu_e\overline{V_c}}{\overline{\sigma_0}}$. This parameter denotes the relative importance of the viscous and surface tension forces. The deformation of the droplets is small and hence this number is kept small. 
    \item Marangoni number given as $Ma = \frac{\beta\overline{GH}}{\mu\overline{V_c}}$. This parameter denotes the relative importance of the temperature-driven Marangoni flow over the strength of the imposed flow. This number is also kept small for the particular problem. 
    \item Thermal Peclet Number given as $Pe = \frac{\overline{V_c}\overline{H}}{\alpha_e}$. this gives the relative importance of the convective transport over the diffusive transport of heat. This is considered small considering the dimensions of the channel. 
    \item Reynolds Number is given as $Re = \frac{\rho_e\overline{V_c}\overline{H}}{\mu_e}$. This shows the relative importance of the convective transport over the diffusive transport of momentum. The Reynolds number is assumed to be small and hence a highly viscous flow is considered. 
    
\end{enumerate}
\subsection{Governing Equations and boundary equations}
The non-dimensional quantities are given as 
\begin{align*}
    x &= \frac{\overline{x}}{\overline{H}} &
    y &= \frac{\overline{y}}{\overline{H}} &
    z &= \frac{\overline{z}}{\overline{H}} & \\
    u_x &= \frac{\overline{u_x}}{\overline{V_c}} &
    u_y &= \frac{\overline{u_y}}{\overline{V_c}} &
    u_z &= \frac{\overline{u_z}}{\overline{V_c}} &
\end{align*}
The non-dimensional velocity distribution in the outside fluid is given as 
\begin{equation}
    (\frac{\partial \vec{u}_e}{\partial t}+(\vec{u_e}.\nabla)\vec{u}_{e})) = -\nabla P_e +\frac{1}{Re}(\nabla^2\vec{u}_e)
\end{equation}
while the non-dimensional velocity distribution in the first droplet is given as 
\begin{equation}
    \rho_r(\frac{\partial \vec{u}_{i1}}{\partial t}+(\vec{u_e}.\nabla)\vec{u}_{i1})) = -\nabla P_i +\frac{1}{Re}(\mu_r\nabla^2\vec{u}_{i1})
\end{equation}
while the non-dimensional velocity distribution in the second droplet is given as 
\begin{equation}
    \rho_r(\frac{\partial \vec{u}_{i2}}{\partial t}+(\vec{u_e}.\nabla)\vec{u}_{i2})) = -\nabla P_i +\frac{1}{Re}(\mu_r\nabla^2\vec{u}_{i2})
\end{equation}
Here $\vec{u}_e$, $\vec{u}_{i1}$, and $\vec{u}_{i2}$ are the non-dimensional velocities of the outside fluid and the first and second droplets respectively. Here $\mu_r = \frac{\mu_i}{\mu_e}$. Here $\mu_r$ is taken to be one. 
While the temperature distribution in the outside fluid is given by
\begin{equation}
    (\frac{\partial T_{e}}{\partial t}+\vec{u_e}. \nabla(T_{e})) = \frac{1}{Pe}\nabla^2 T_{e}
\end{equation}
The temperature distribution in the first droplet is given as 
\begin{equation}
    \rho_{r}c_{r}(\frac{\partial T_{i1}}{\partial t}+(\vec{u_i}.\nabla)T_{i1}) =\frac{1}{Pe}\nabla^2 T_{i1}
\end{equation}
The temperature distribution in the second droplet is given as 
\begin{equation}
    \rho_{r}c_{r}(\frac{\partial T_{i2}}{\partial t}+(\vec{u_i}.\nabla)T_{i2}) =\frac{1}{Pe}\nabla^2 T_{i2}
\end{equation}
Here $T_{e}$, $T_{i1}$, and $T_{i2}$ are the non-dimensional temperatures of the fluid and droplet respectively. Here $\rho_r = \frac{\rho_i}{\rho_e}$ 
\par The Reynolds numbers and Peclet numbers are assumed to be low so the previous equation can be modified to be as (assuming steady state). Thus we have the energy equation in the fluid as
\begin{equation}\label{heat_eq_1}
   k_{e}\nabla^2 T_{e} = 0
\end{equation}
For the 1st droplet, we have  
\begin{equation}\label{heat_eq_2}
   k_{i}\nabla^2 T_{i1} = 0
\end{equation}
For the droplet we have 
\begin{equation}\label{heat_eq_3}
   k_{i}\nabla^2 T_{i2} = 0
\end{equation}
The boundary conditions at the interface of the droplets and main fluid are  
\begin{equation}\label{heat_bd_1}
    T_{i1} = T_{e} = T_{s1}
\end{equation} and 
\begin{equation}\label{heat_bd_2}
    \delta_2(\hat{n}.\nabla T_{i1}) = \hat{n}.\nabla T_{e}
\end{equation} 
for the second droplet, we have 
\begin{equation}\label{heat_bd_3}
    T_{i2} = T_{e} = T_{s2}
\end{equation} and 
\begin{equation}\label{heat_bd_4}
    \delta(\hat{n}.\nabla T_{i2}) = \hat{n}.\nabla T_{e}
\end{equation} 
Here $\delta = \frac{k_i}{k_e}$ is the conductivity ratio between the fluid inside the droplet and the continuous medium. 
The boundary conditions at the inlet and outlet are given as 
%\begin{multline}\label{heat_bd_5}
\begin{align}
    T_{i1} = T_{i2} = T_{\infty} = \gamma z \label{heat_bd_5}\\
    T_{e} = T_{\infty} = \gamma z\label{heat_bd_6}
\end{align}
%\end{multline}
The walls are considered to be insulated
\begin{equation}\label{heat_bd_7}
    \hat{n}.\nabla T_W = 0
\end{equation}
For the fluid flow, a steady Stokes equation is considered since Reynolds numbers are assumed to be low which indicates the flow reaches a Stokes flow form given by 
\begin{equation}
    \nabla P_e = \nabla^2 u_e.
\end{equation}
For the fluid medium inside the 1st droplet, we have
\begin{equation}
    \nabla P_{i1} = \nabla^2 u_{i1}.
\end{equation}
For the second droplet, we have 
\begin{equation}
    \nabla P_{i2} = \nabla^2 u_{i2}.
\end{equation}
The continuity equations are given as
\begin{equation}
    \nabla. u_{e} = 0
\end{equation}
\begin{equation}
    \nabla. u_{i1} = 0
\end{equation}
\begin{equation}
    \nabla. u_{i2} = 0
\end{equation}
The imposed velocity profile is given as 
\begin{equation}\label{vel_inf}
\vec{u_{\infty}} = (1-4\frac{z^2}{H^2})\hat{x}
\end{equation}
The boundary conditions at the walls is no slip which means 
that at the walls
\begin{equation}\label{fluid_bd_1}
 \vec{u_e} = 0   
\end{equation}
Here the boundary conditions at the interface between the outer fluid and the droplet are 
\begin{equation}\label{fluid_bd_2}
\vec{u}_e.\hat{n} = \vec{u}_{i1}.\hat{n} = \frac{dx_{s1}}{dt}
\end{equation}
\begin{equation}\label{fluid_bd_3}
\vec{u}_e - (\vec{u}_e.\hat{n}).\hat{n}=\vec{u}_{i1} - (\vec{u}_{i1}.\hat{n}).\hat{n}
\end{equation}
For the second droplet, we have 
\begin{equation}\label{fluid_bd_4}
\vec{u}_e.\hat{n} = \vec{u}_{i2}.\hat{n} = \frac{dx_{s2}}{dt}
\end{equation}
\begin{equation}\label{fluid_bd_5}
\vec{u}_e - (\vec{u}_e.\hat{n}).\hat{n}=\vec{u}_{i2} - (\vec{u}_{i2}.\hat{n}).\hat{n}
\end{equation}
The traction equations at the interface of the fluid and the droplet. 
\begin{equation}\label{fluid_bd_6}
(S_e-S_{i1}).\hat{n} = \sigma_{1}\hat{n}(\nabla.\hat{n})-\nabla_S\sigma_{1}
\end{equation}
\begin{equation}\label{fluid_bd_7}
(S_e-S_{i2}).\hat{n} = \sigma_{2}\hat{n}(\nabla.\hat{n})-\nabla_S\sigma_{2}
\end{equation}
The vectors $\vec{u}_e$, $\vec{u}_{i1}$ and $\vec{u}_{i2}$ are the flow velocities outside and inside the droplet 1 and droplet 2 while $S_e$, $S_{i1}$ and $S_{i2}$ are the corresponding stresses. The term $\sigma_{1}$ is the surface tension which is a linear function of the surface temperature given as
\begin{equation}\label{fluid_bd_8}
\sigma_{1}= \sigma_0-\beta(T_{s1}-T_0)
\end{equation}
\begin{equation}\label{fluid_bd_9}
\sigma_{2}= \sigma_0-\beta(T_{s2}-T_0)
\end{equation}
Here $s\sigma_0$ is the surface tension at temperature $T_0$ while the term $\beta$ is given as $\beta = d\sigma/dT >0$. The temperatures $T_{s1}$ and $T_{s2}$ are the temperatures at the interface of the first and second droplets respectively. 
\par The evolution of the droplet surface is given by the following equations that give 
\begin{equation}\label{fluid_bd_10}
    \frac{dx_{s1}}{dt} .\hat{n} = \vec{u_{i1}}.\hat{n} = \vec{u_{e}}.\hat{n} 
\end{equation}
\begin{equation}\label{fluid_bd_11}
    \frac{dx_{s2}}{dt} .\hat{n} = \vec{u_{i2}}.\hat{n} = \vec{u_{e}}.\hat{n} 
\end{equation}
The above equation can be written ignoring any phase change phenomenon happening at the interface of the droplets. The normal velocity at the interface contributes to the droplet interface evolution since the droplet phase change occurs due to the normal velocity while the tangential velocity assists in translating the droplets. 
\section{Methodology}
\subsection{Numerical Methodology}
\subsubsection{Energy equation formulation}
The \cref{heat_eq_1,heat_eq_2,heat_eq_3} along with the boundary conditions \cref{heat_bd_1,heat_bd_2,heat_bd_3,heat_bd_4,heat_bd_5,heat_bd_6,heat_bd_7} are solved by the boundary element method. The boundary element method is chosen because of the lack of need for volume discretization and this generates a grid on the surface that reduces the number of computational points. The boundary integral equations formed from the equations are given below. For the sake of brevity, the full derivation of the boundary integral forms has been omitted. The terms $T_{s1}$,$T_{s2}$,$T_w$, represent the temperatures on the surface of the first and second droplet and the wall respectively. 
equation is given for the collocation point on the surface of the first droplet 
\begin{multline}\label{temp_1}
\frac{T_{s1}(x_0)(1+\delta)}{2} = T_{\infty}+ (1-\delta)\int_{s1}^{PV}((\hat{n}.\nabla G(x,x_0))T_{s1} ds)\\+(1-\delta)\int_{s2}((\hat{n}.\nabla G(x,x_0))T_{s2} ds)+\int_W (\hat{n}.\nabla G(x,x_0)T_w.ds)
\end{multline}
For the collocation point on the second droplet surface 
\begin{multline}\label{temp_2}
\frac{T_{s2}(x_0)(1+\delta)}{2} = T_{\infty}+ (1-\delta)\int_{s2}^{PV}((\hat{n}.\nabla G(x,x_0))T_{s2} ds)\\+(1-\delta)\int_{s1}((\hat{n}.\nabla G(x,x_0))T_{s1} ds)+\int_W (\hat{n}.\nabla G(x,x_0)T_w.ds)
\end{multline}
For the collocation point on the surface of the wall
\begin{multline}\label{temp_3}
\frac{T_{w}(x_0)}{2} = T_{\infty}+ (1-\delta)\int_{s2}((\hat{n}.\nabla G(x,x_0))T_{s2} ds)\\+(1-\delta)\int_{s1}((\hat{n}.\nabla G(x,x_0))T_{s1} ds)+\int_W^{PV} (\hat{n}.\nabla G(x,x_0)T_w.ds)
\end{multline}
These equations \cref{temp_1,temp_2,temp_3} can be used to solve for the temperatures($T_{s1}$,$T_{s2}$ and $T_w$). 
\subsubsection{Fluid formulation}
For the velocity,  we have the integral formulation for flow across an interface for two liquids of equal viscosity. This formulation is for points on the interface of the droplet and the main fluid. 
For the first droplet, we have the following velocity formulation,
\begin{multline}\label{vel_1}
\vec{u_{i1}}(x_0) = \vec{u}_{\infty}(x_0)-\frac{1}{8\pi\mu}\int^{PV}_{s1}((S_e-S_{i1}).\hat{n})[G]ds\\-\frac{1}{8\pi\mu}\int_{s2}((S_e-S_{i2}).\hat{n})[G]ds 
-\frac{1}{8\pi\mu}\int_{W}(S_w.\hat{n})[G]ds
\end{multline}
For the second droplet, we have the following 
\begin{multline}\label{vel_2}
\vec{u_{i2}}(x_0) = \vec{u}_{\infty}(x_0)-\frac{1}{8\pi\mu}\int^{PV}_{s2}((S_e-S_{i2}).\hat{n})[G]ds\\-\frac{1}{8\pi\mu}\int_{s1}((S_e-S_{i1}).\hat{n})[G]ds 
-\frac{1}{8\pi\mu}\int_{W}(S_w.\hat{n})[G]ds
\end{multline}
Here the boundary conditions at the interface are given by \cref{fluid_bd_1,fluid_bd_2,fluid_bd_3,fluid_bd_4,fluid_bd_5,fluid_bd_6,fluid_bd_7,fluid_bd_8,fluid_bd_9,fluid_bd_10,fluid_bd_11}. The boundary conditions are used to solve for the unknown wall traction term. 
To calculate the surface velocity of the droplet we have to use the below two equations for when the collocation point is at the wall and droplet surface respectively. 
\begin{multline}\label{vel_3}
0 = \vec{u}_{\infty}(x_0)-\frac{1}{8\pi\mu}\int_{s2}((S_e-S_{i2}).\hat{n})[G]ds\\-\frac{1}{8\pi\mu}\int_{s1}((S_e-S_{i1}).\hat{n})[G]ds 
-\frac{1}{8\pi\mu}\int^{PV}_{W}(S_w.\hat{n})[G]ds
\end{multline}
 From \cref{fluid_bd_1}the free stream velocity at the wall is zero ($u_{\infty}$). This means that \cref{vel_3} reduces to 
\begin{multline}\label{vel_4}
-\frac{1}{8\pi\mu}\int_{s2}((S_e-S_{i2}).\hat{n})[G]ds\\-\frac{1}{8\pi\mu}\int_{s1}((S_e-S_{i1}).\hat{n})[G]ds 
 = \frac{1}{8\pi\mu}\int^{PV}_{W}(S_w.\hat{n})[G]ds
\end{multline}
From equation \cref{vel_4} the unknown term $(S_{w}.\hat{n})$ can be calculated. 
The computed temperature $T_s$ is used to calculate the surface tension and stress variation across the interface from \cref{fluid_bd_6,fluid_bd_7}. Then using \cref{vel_1,vel_2} the velocity at the interface of the droplet ($\vec{u_{i1}}$) and ($\vec{u_{i2}}$) can be calculated.  
\subsubsection{Interface Tracking}
The interface is tracked using a second order runge kutta scheme given as 
\begin{equation}
    x_{n+1} = x_n+\frac{\delta t}{2}(u_n+u_{n+\frac{1}{2}})
\end{equation}
Here the terms $u_n$ and $u_{n+\frac{1}{2}}$ represent the velocities of the marker points at the nth time step and the intermediate time step computed from the values of the position and velocity of the interface at the n-th step. The velocity $u_n$ is the superposition of the normal velocity at the interface marker points and the mesh relaxation velocity. This Mesh relaxation velocity is added to ensure stability of the mesh at high levels of deformation and does not lead to any change in the physics of the problem. 
\section{Results and discussion}
\subsection{Validation}
\begin{figure}[H]
\centering
    \includegraphics[width=0.5\textwidth]{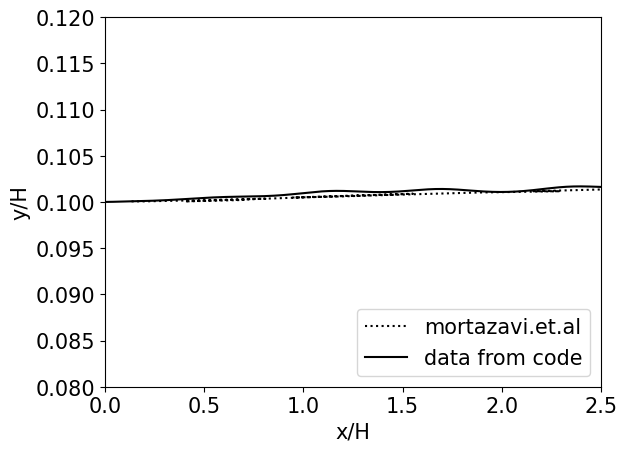}
     \caption{Vertical movement of the droplet in the absence of temp gradient}
     \label{fig:figure1}
\end{figure}

\begin{figure}[!htb]
     \centering
     \includegraphics[width=0.5\textwidth]{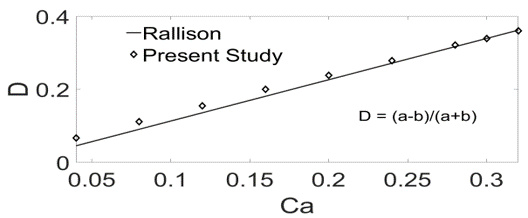}
     \caption{Droplet deformation with capillary number}
     \label{fig:figure7}
\end{figure}

To assess the method's accuracy, we considered the migration of a single droplet in the presence of a Posieulle flow bounded by no-slip walls. We aim to validate our results with those of \citet{Mortaza2002}. 
The flow is assumed to be steady, incompressible, and in the low Reynolds number regime. The droplet viscosity is considered to be the same as that of the bulk fluid. The capillary number drives the flow absence of the Reynolds number in the Stokes limit and is defined as \begin{equation*}
    Ca=\frac{U_{0}\mu_{o}}{\sigma}
\end{equation*}
The non-dimensional droplet radius is taken to be 0.125. The viscosity of both droplet and fluid is taken to be the same. The capillary number has been taken to be 0.25. 
%The no-slip conditions have been applied at the wall and the periodic boundary conditions at the inlet and the outlet with the inlet flow being fully developed poiseuille flow.
The droplet is positioned at a distance above the centreline of the channel which is taken to be 0.1. In the first figure, we see the migration of the drop in the absence of a thermal gradient. The locus of the droplet centroid is compared with those of \cite{Mortaza2002} as depicted in \cref{fig:figure1}. The overall error is of the order of around less than one percent.  
It can also be inferred that for the single droplet there are only two forces acting one of the wall-induced forces and those of the Marangoni forces. For low diameters we have the droplet moving towards the wall while for the higher diameters the droplet moves downward towards the centre. 
% The sets of results corroborate well with those of \citet{Mortaza2002}.
For the case of multiple droplets, we consider the case of the droplets moving in uniform shear flow. The droplet deformation is plotted against the capillary number. The deformation of the drops is plotted with the Capillary number. The results are shown in \cref{fig:figure7}
% \subsection{Migration of two drops in the absence of an temperature gradient}
% \subsubsection{Effect of initial droplet center separation and confinement ratio}
% The overall thermocapillary migration is governed by three forces namely the wall-induced lift forces and the marangoni forces due to the temperature gradient as described in \citet{das2018}. Understanding the nature of these forces is vital to characterizing the motion of a droplet system in a confined channel. These forces depend on a variety of factors which include the droplet confinement ratio, the droplet separation distances, the overall geometry of the channel, and the magnitude of the temperature gradient. Some of the parameters responsible for the variation in these forces is Capillary number($Ca$), Marangoni number($Ma$), confinement ratio($Cr$) and droplet separation distance($d-d_{0}$). It can be assumed that the wall induced lift forces and the droplet interaction forces are not dependent on the Marangoni number($Ma$) while the thermocapillary forces are dependent on the Marangoni number($Ma$). In this regard, we look at the system in the absence of any thermocapillary forces. The nature of the droplet interaction forces also varies from one droplet to another which means that these forces act differently on the second droplet. 
\subsection{Migration of two drops in the presence of an axial temperature gradient}
\subsubsection{Effect of confinement ratio on the droplet migration and droplet separation distance}
The overall thermocapillary migration is governed by certain forces namely the hydrodynamic forces due to the droplet-droplet and droplet-wall interactions and the thermal forces due to the temperature gradient as described in \citet{das2018}. Understanding the nature of these forces is vital to characterizing the motion of a system of droplets in a confined channel. These forces depend on a variety of factors which include the droplet confinement ratio, the droplet separation distances, the overall geometry of the channel and the magnitude of the temperature gradient. Some of the parameters responsible for the variation in these forces are the capillary number ($Ca$), the Marangoni number ($Ma$), the confinement ratio ($Cr$) and the initial droplet separation distance($d-d_{0}$). Intuitively, it can be stated that the confinement ratio($Cr$) affect the droplet-wall forces and droplet-droplet interaction forces, while the droplet-droplet interaction forces are also dependent on the droplet separation distance($d-d_{0}$). The thermal forces are influenced by the marangoni number($Ma$). In a microfluidic system, experiments have shown that thermal Marangoni numbers can be varied from 1-5. The capillary numbers are kept small so as not to consider the effects of large droplet deformation. 
\par
\begin{figure}[!htb]
    \centering
    \includegraphics[width=0.5\textwidth]{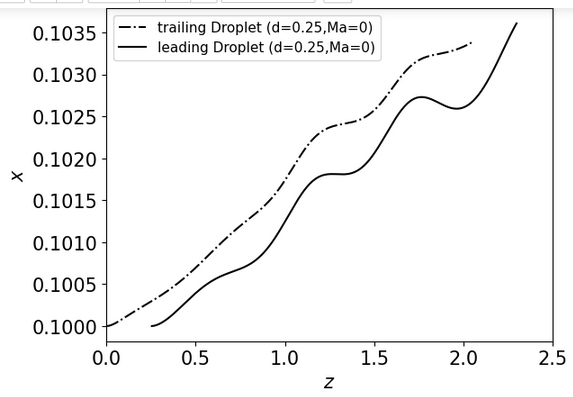}
    \vspace{-0.1in}
    \caption{Droplet Migration of two drops in contact in the absence of temperature gradient ($a = 0.125$)}
    \label{fig:figure71}
\end{figure}

 Two equiviscous droplets of equal radius are positioned such that they touch each other axially along the lines of their centers to study their dynamics in the presence and absence of an imposed axially oriented temperature gradient. Drops are placed at an offset distance ($e=p=0.1H$) and their migration characteristics are studied by changing their droplet separation($d$) and radius($a$). The trailing droplet is placed at the channel inlet while the leading droplet is translated in the x-direction by an amount equal to the separation distance($d$). The computational domain is shown in \cref{main_pic}. The capillary number($Ca$) is taken to be 0.25. In our first study, an isothermal flow is assumed to assess the nature of the droplet-droplet interaction and droplet-wall interaction forces. The droplet separation($d$) is taken as 0.25 while the droplet radius($a$) is taken as 0.125. The droplet migration trajectories in the absence of a temperature gradient are shown in \cref{fig:figure71}. 
\par
\begin{figure}[!htb]
\centering
\begin{subfigure}{0.5\textwidth}
\centering
\includegraphics[width=0.7\textwidth]{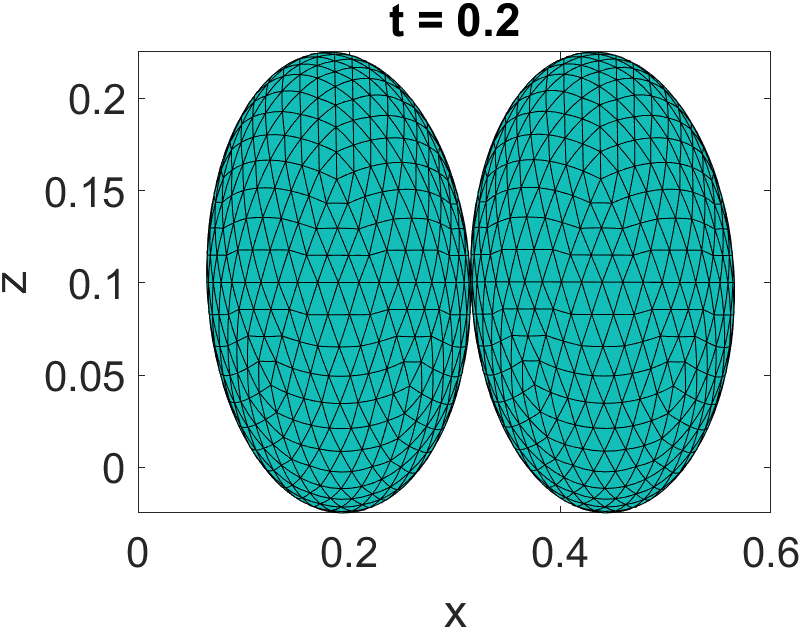}  
\end{subfigure}%
\begin{subfigure}{0.5\textwidth}
\centering
\includegraphics[width=0.7\textwidth]{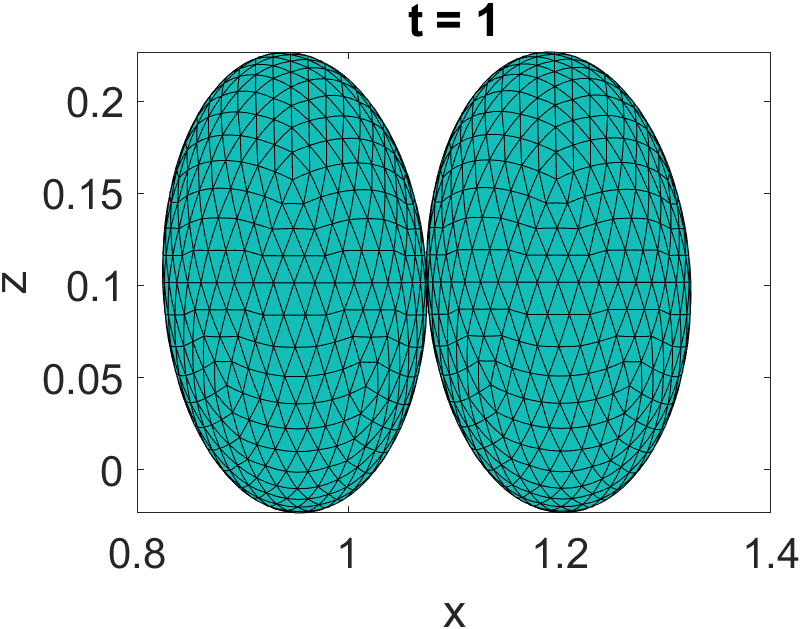}  
\end{subfigure}
\caption{Depiction of the migrating droplets}
\label{fig:fig73}
\end{figure}

 In the absence of the temperature gradient a transverse droplet migration towards the wall is observed as has been observed previously by \cite{Mortaza2002} for the case of a single droplet. It is also observed that in the quasi-steady limit in the low Reynolds regime, the Peclet numbers, along with isothermal conditions, flow droplets of equal size as conjoined droplets and flow together as one unit as depicted in \cref{fig:fig73}. Both the droplet-droplet interaction forces and wall forces tend to lift the droplet upward towards the wall in the absence of an imposed temperature field. 
% The forces acting on each droplet in this case, are the the hydrodynamic lift force due to the droplet deformation and the droplet interaction forces. These result in the upward migration of the droplets. We compare this with the behavior of single droplets and see that the upward migration rate of the multiple droplets is more pronounced. This has been described in \cref{fig:figure72p}. This shows that in the presence of the droplet interaction forces, the droplet transverse migration towards the wall is more rapid thus it can be inferred that the effect of the droplet interaction forces is to push the droplet towards the wall. 
% \begin{figure}[!htb]
%     \centering
%     \includegraphics[width=0.5\textwidth]{single_multiple_0.125_0.25.png}
%     \vspace{-0.1in}
%     \noindent\caption{Comparison of single and multiple droplet trajectories}
%     \label{fig:figure72p}
% \end{figure}
\par
\begin{figure}[H]
    \centering
    \includegraphics[width=0.5\textwidth]{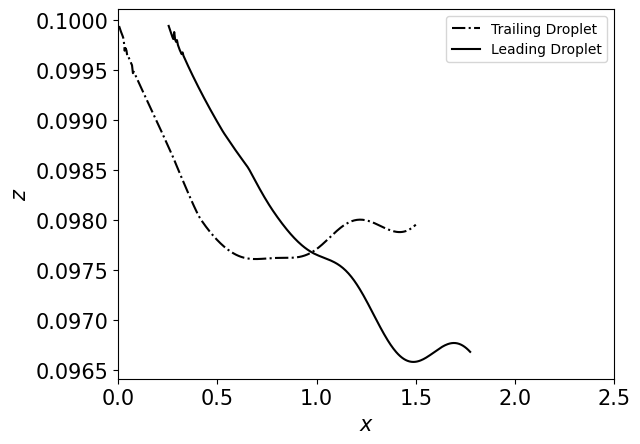}
    \vspace{-0.1in}
    \noindent\caption{Droplet Migration of two drops in contact in the presence of temperature gradient($Ma = 0.5$)}
    \label{fig:figure72}
\end{figure}

\begin{figure}[!htb]
\centering
\begin{subfigure}{0.5\textwidth}
\centering
\includegraphics[width=0.7\textwidth]{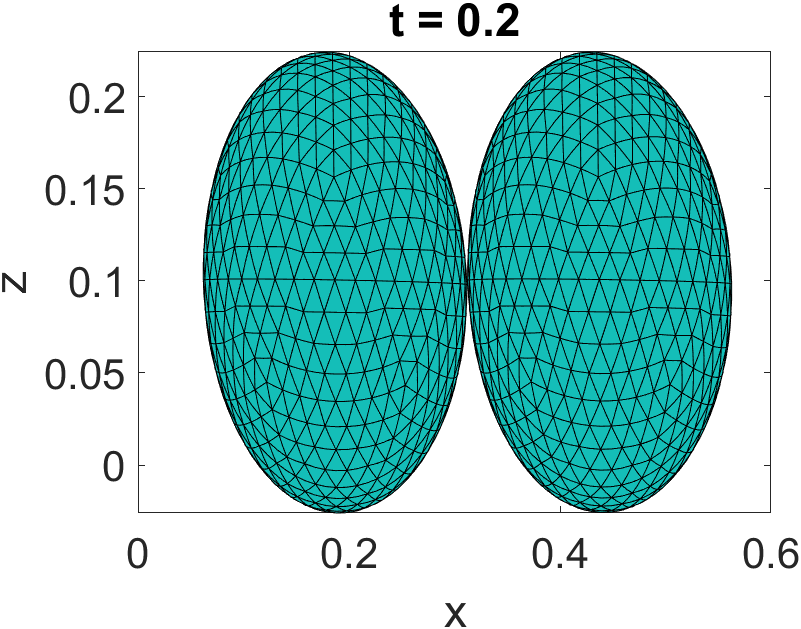}  
\end{subfigure}%
\begin{subfigure}{0.5\textwidth}
\centering
\includegraphics[width=0.7\textwidth]{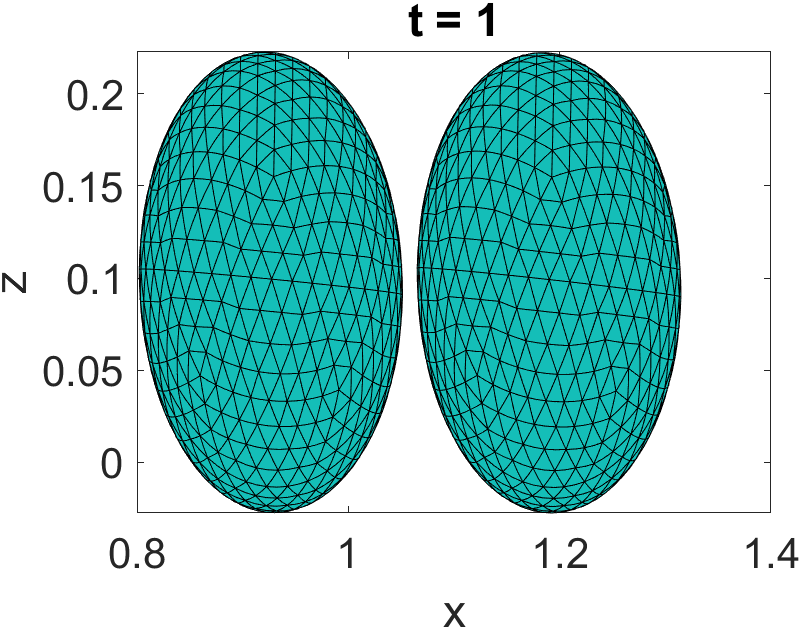}  
\end{subfigure}
\caption{Depiction of the migrating droplets}
\label{fig:fig74}
\end{figure}

\begin{figure}[H]
     \centering
     \includegraphics[width=0.5\textwidth]{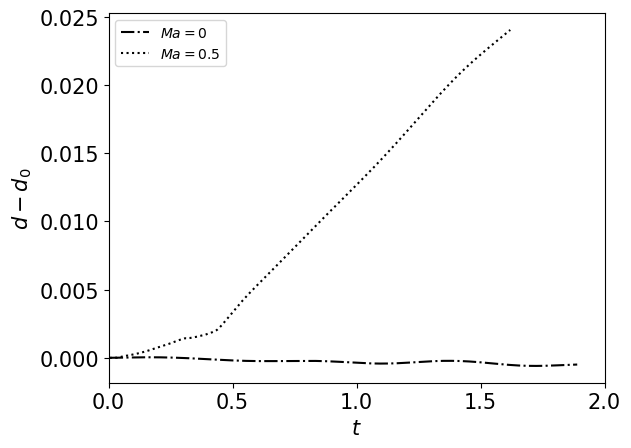}
     \caption{Droplet separation distances with the time scale}
     \label{fig:figure8}
 \end{figure}

\begin{figure}[!htb]
\centering
    \begin{subfigure}[t]{0.5\textwidth}
    \centering
    \includegraphics[width=0.9\textwidth]{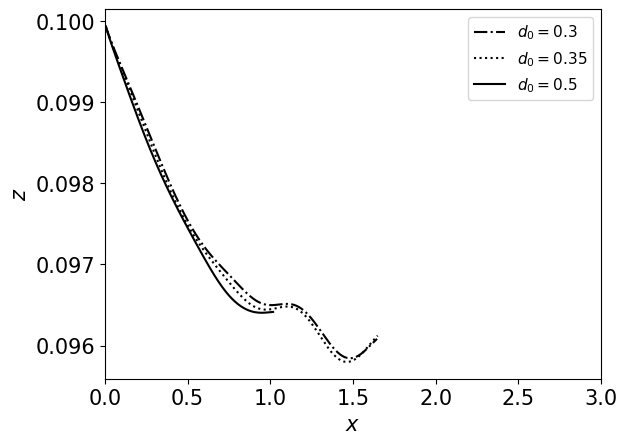}
    \caption{Trailing Droplet}
    \label{fig:figure18}  
    \end{subfigure}%
    \begin{subfigure}[t]{0.5\textwidth}
    \centering
    \includegraphics[width=0.9\textwidth]{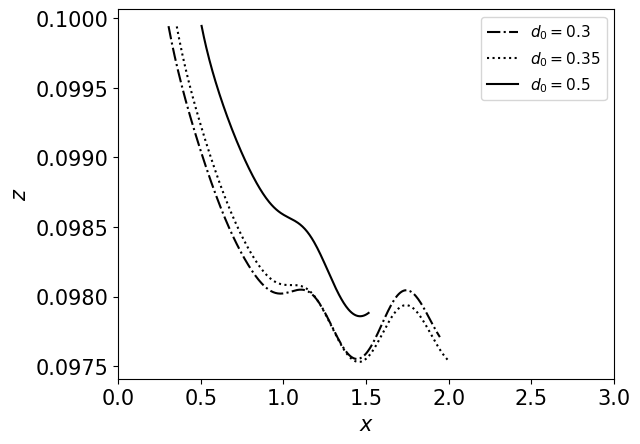}
    \caption{Leading Droplet}
    \label{fig:figure19} 
    \end{subfigure}
    \caption{Droplet Migration trajectories ($a = 0.125$)}
    \label{fig:figure9}
\end{figure}

\begin{figure}[!htb]
    \centering
    \includegraphics[width=0.5\textwidth]{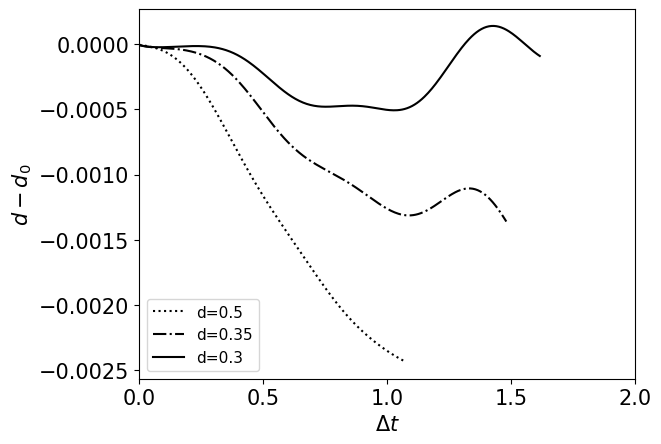}
    \caption{Droplet separation distances with time scale ($a = 0.125$)}
    \label{fig:figure22}
\end{figure}

\begin{figure}[!htb]
\noindent
\begin{subfigure}{1.0\textwidth}
\centering
\begin{tabular}{|c|c|c|c|}
      \hline
      \addheight{\includegraphics[width=30mm]{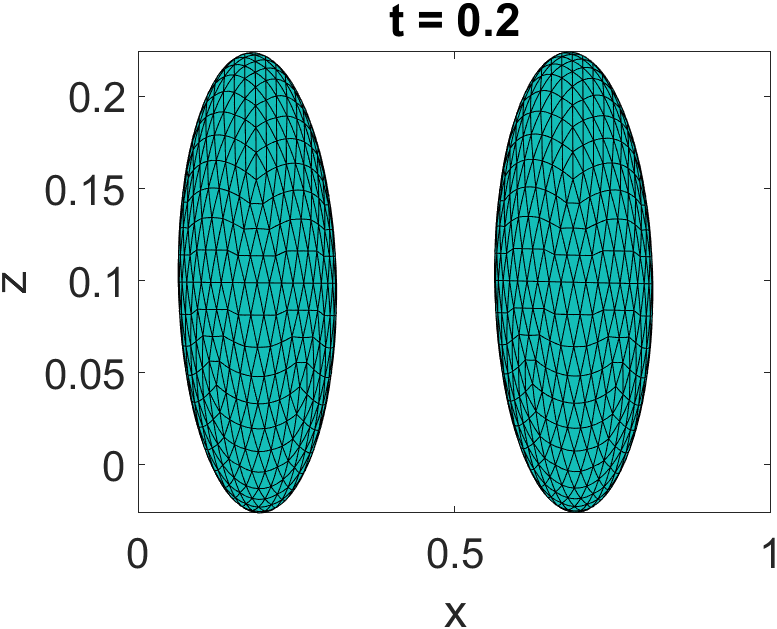}} &
      \addheight{\includegraphics[width=30mm]{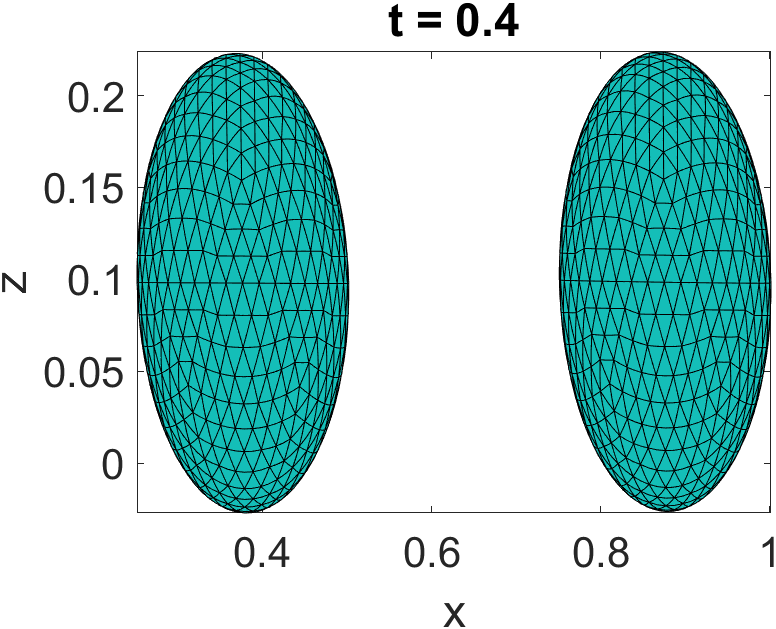}} &
      \addheight{\includegraphics[width=30mm]{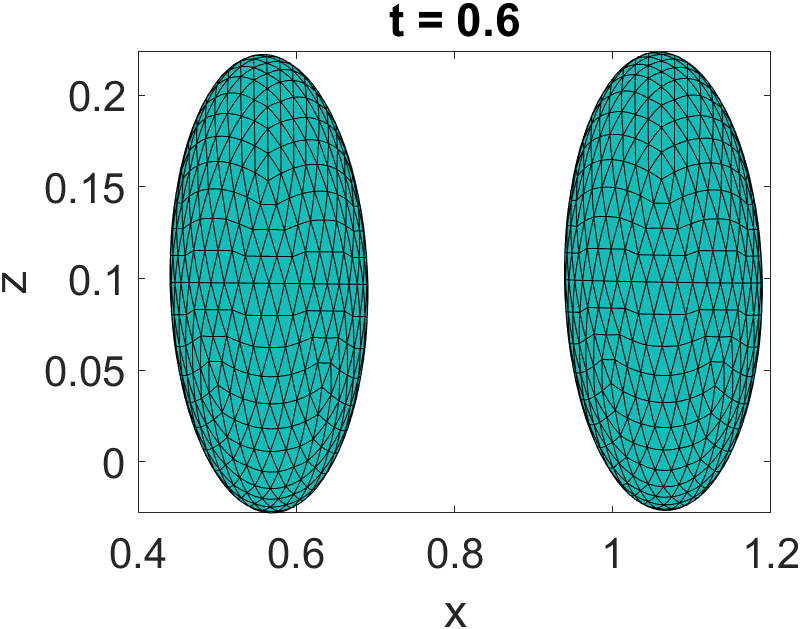}} &
      \addheight{\includegraphics[width=30mm]{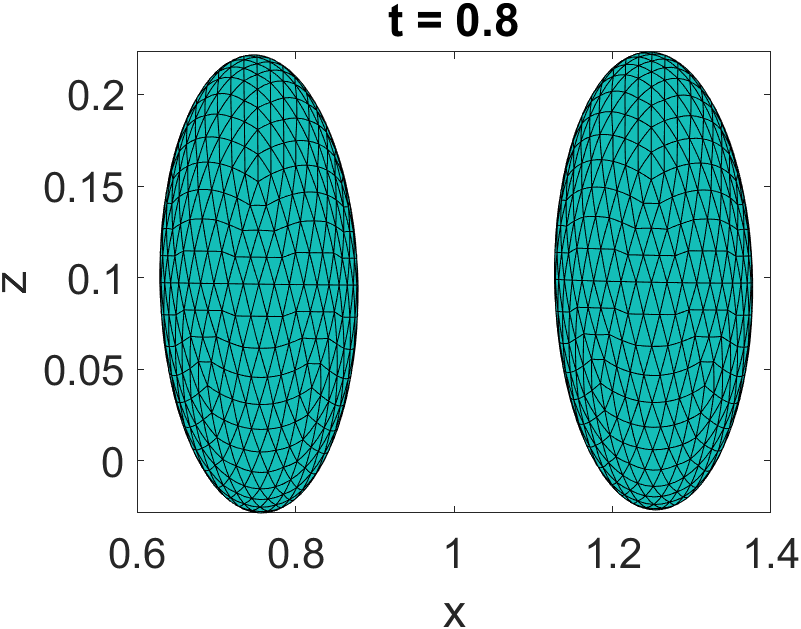}} \\
      \hline     
\end{tabular}
% \caption{evolution of droplet shapes ($a = 0.125,d = 0.5)$}
% \label{fig:figure83}
\end{subfigure}
\begin{subfigure}{1.0\textwidth}
\centering
\begin{tabular}{|c|c|c|c|}
      \hline
      \addheight{\includegraphics[width=30mm]{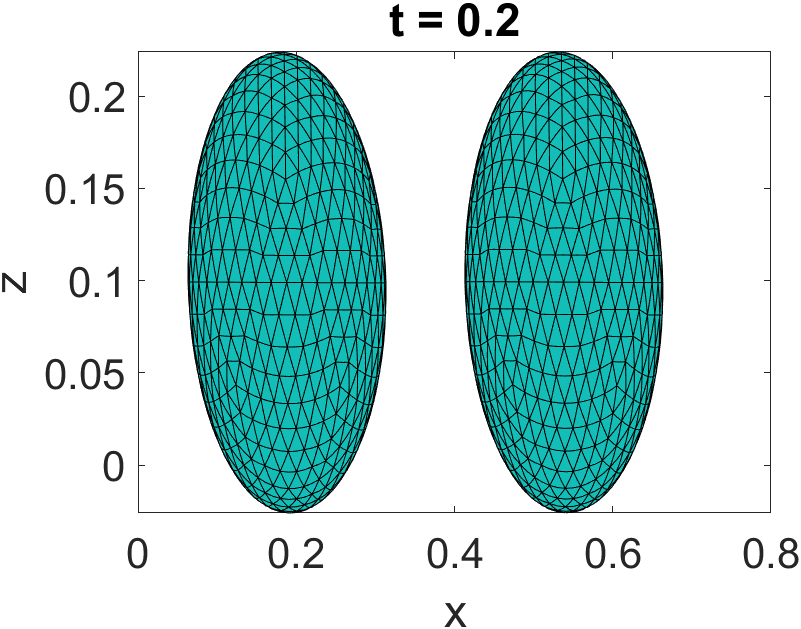}} &
      \addheight{\includegraphics[width=30mm]{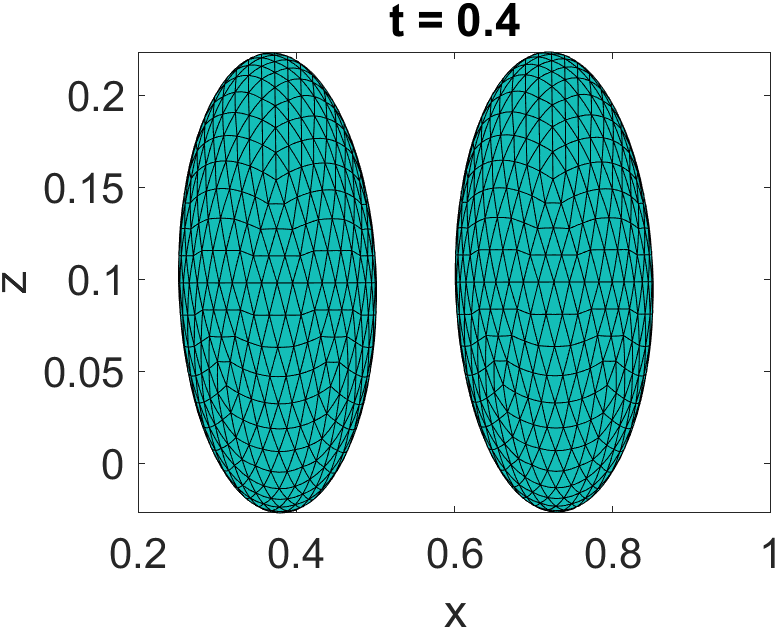}} &
      \addheight{\includegraphics[width=30mm]{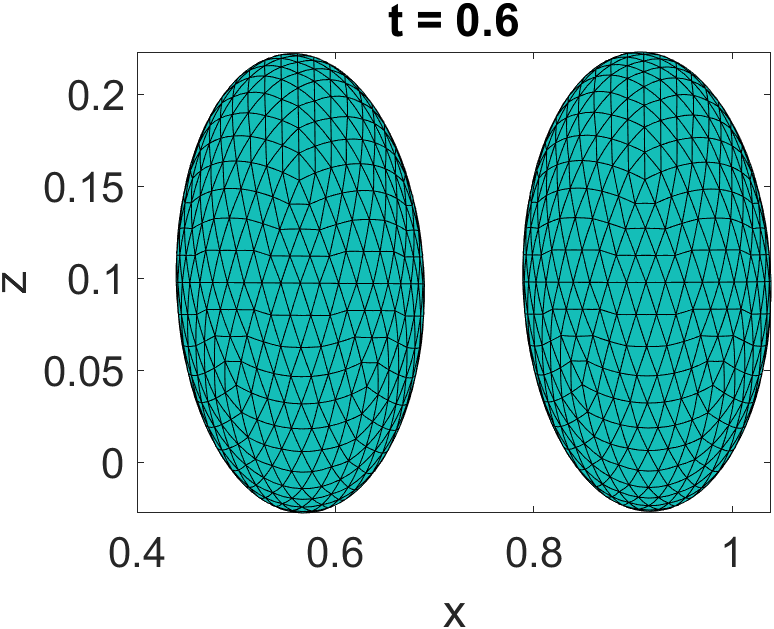}} &
      \addheight{\includegraphics[width=30mm]{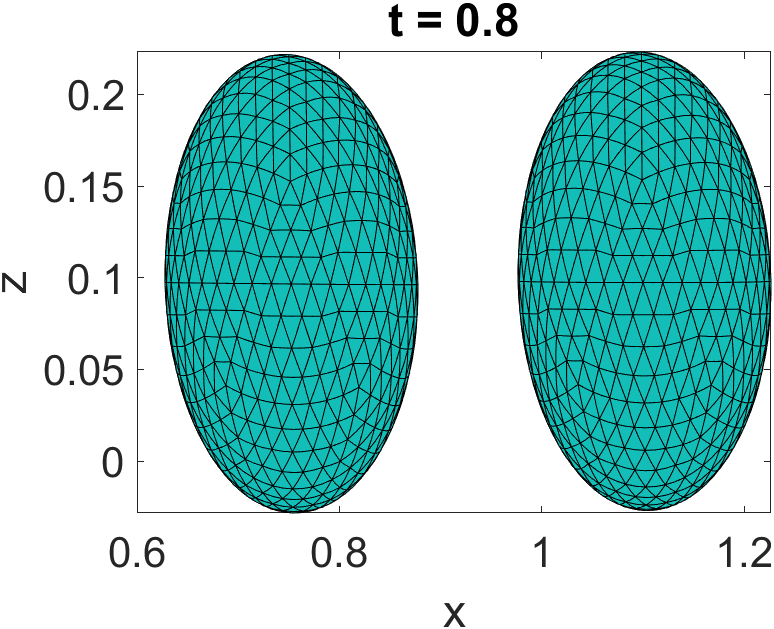}} \\
      \hline     
\end{tabular}
% \caption{evolution of droplet shapes ($a = 0.125,d = 0.5)$}
% \label{fig:figure83}
\end{subfigure}
\begin{subfigure}{1.0\textwidth}
\centering
\begin{tabular}{|c|c|c|c|}
      \hline
      \addheight{\includegraphics[width=30mm]{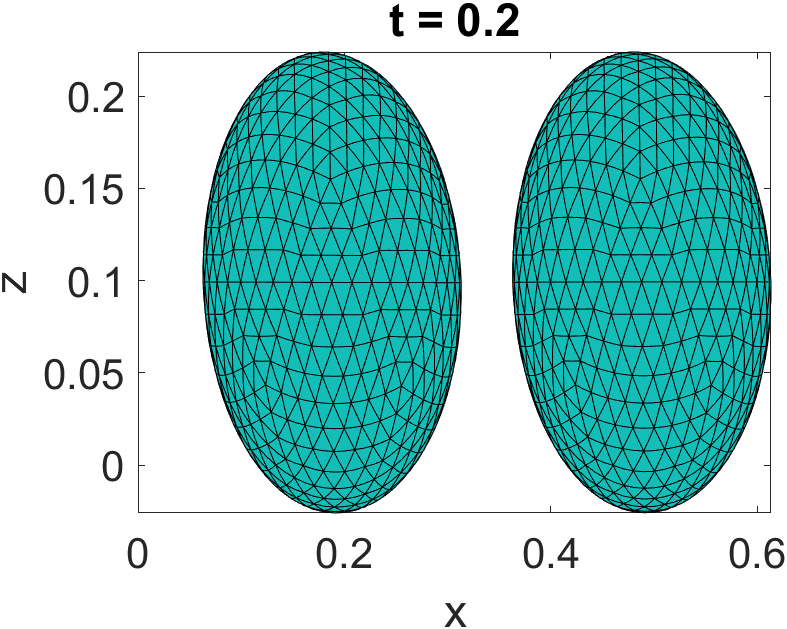}} &
      \addheight{\includegraphics[width=30mm]{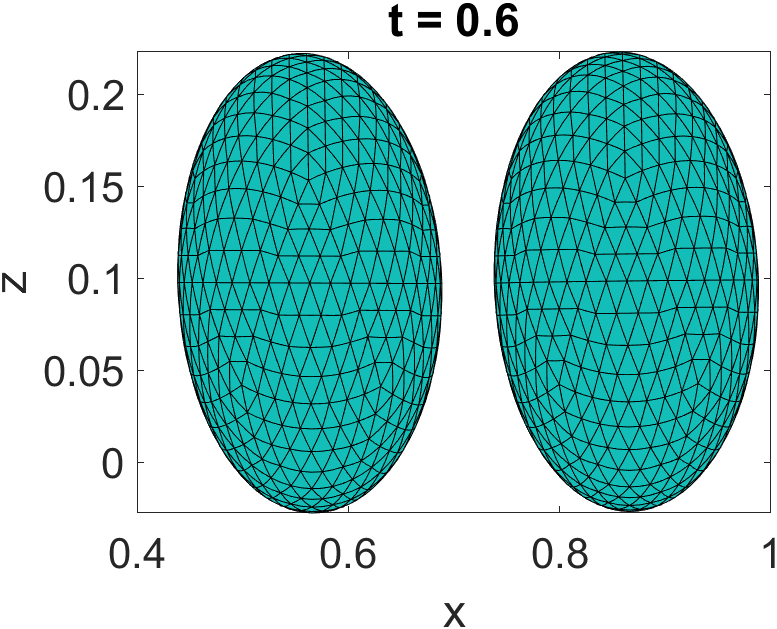}} &
      \addheight{\includegraphics[width=30mm]{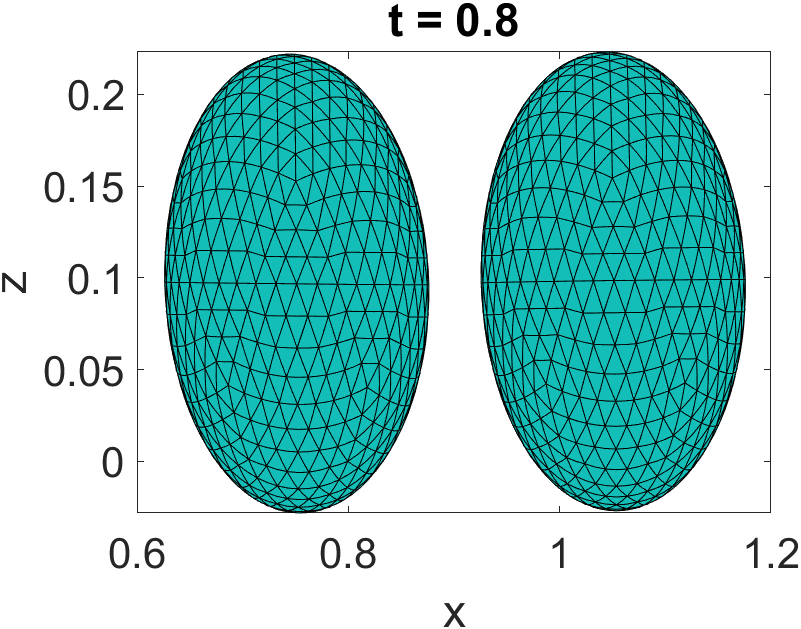}} &
      \addheight{\includegraphics[width=30mm]{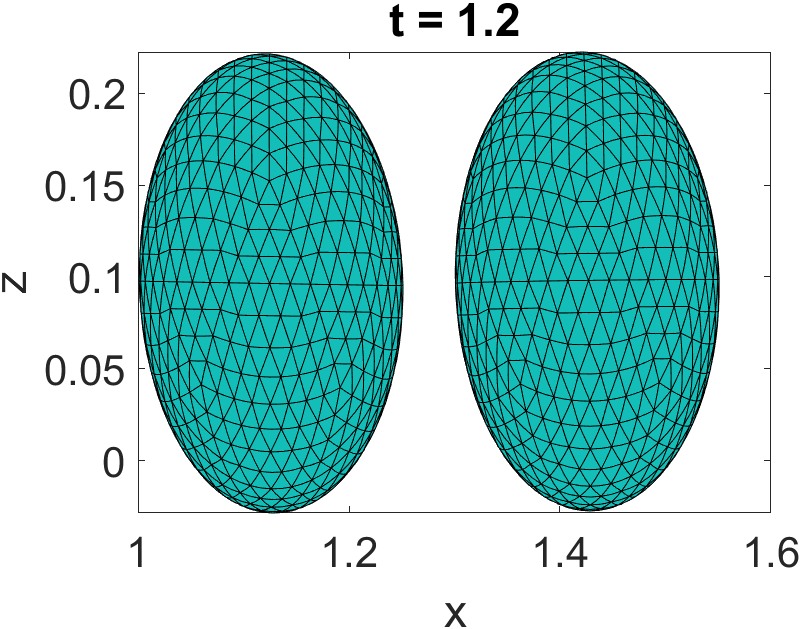}} \\
      \hline     
\end{tabular}
% \caption{evolution of droplet shapes ($a = 0.125,d = 0.5)$}
% \label{fig:figure83}
\end{subfigure}
\caption{Evolution of droplet shapes ($a = 0.125$): Top row ($d_{0} = 0.5$) Middle row ($d_{0} = 0.35$) Bottom row ($d_{0} = 0.3$)}
\label{fig:figure831}
\end{figure}

 Next, we consider the effect of imposing an axial temperature gradient on the flow increasing along the positive x-axis. The Marangoni number of the flow is taken to be 0.5. The droplet migration is observed in the opposite transverse direction that leads to the droplet migrating to the centreline as described in \cref{fig:figure72}. 
The migration trajectories of the droplets away from the walls and towards the channel centreline had been observed previously in the case of single drops within insulated walls by \citet{capobianchi2017}. The thermal forces overpower the droplet-wall and droplet-droplet interaction forces and cause downward migration of the droplet and drives the droplet towards the centreline. Here, the two drops remain conjoined at first and then gradually separate as depicted in \cref{fig:fig74}. This is due to the axially varying temperature gradient, which results in a different force acting on the leading droplet compared to that of the trailing drop that causes their separation. To demonstrate the conditions under which the drops are separating from one another, the differences in the droplet center distances are plotted with the change in the time for the cases of the temperature gradient imposed and not imposed. The effect is visualized in \cref{fig:figure8}.
 Now we have to see how the behavior in droplets changes with confinement ratio and separation distance. Previous parameters like ($e=p=0.1H$) and ($Ca=0.25, Ma=0.5$) are kept unchanged while the initial separation of the droplets($d_{0}$) is altered. We see how the migration trajectories of the drops are affected by initial droplet spacing for a particular value of the confinement ratio. Initial simulations are performed for a droplet radius($a$) of $0.125H$ and various initial separation distances($d_{0}$) of $0.3H, 0.35H$, and $0.5H$. The trailing droplets' migration trajectories are observed and plotted in \cref{fig:figure9}. In the case of the trailing droplet the migration trajectories of the drops are approximately the same as the droplet descent however the rate of descent of the droplet becomes higher for the larger initial separation of the droplets. Due to the higher separation between the droplets, the droplet interaction forces that push the droplets toward the wall are lower and their effects are reduced in the face of the higher thermal forces. 
% \par It can be seen from \cref{fig:figure10} that the transverse migration velocity of the trailing droplet increases with droplet separation while for the leading droplet, the migration velocity decreases with droplet separation. Overall for any degree of separation, the transverse migration velocity of the trailing droplet first increases, then decreases while for the leading droplet, this migration velocity remains the same. The transverse migration velocity of the trailing droplet reduces as it nears the leading droplet. 
% Hence we see overall lower velocities for the trailing droplets at smaller values of initial droplet separation. 
% \begin{figure}[H]
% \centering
%     \begin{subfigure}[t]{0.5\textwidth}
%     \centering
%     \includegraphics[width=0.9\textwidth]{d_0.35_0.5_0.3_velz.png}
%     \noindent\caption{Trailing Droplet}
%     \label{fig:figure20}  
%     \end{subfigure}%
%     \begin{subfigure}[t]{0.5\textwidth}
%     \centering
%     \includegraphics[width=0.9\textwidth]{d_0.35_0.5_0.3_leading_velz.png}
%     \noindent\caption{Leading Droplet}
%     \label{fig:figure21} 
%     \end{subfigure}
%     \caption{Droplet migration transverse velocities($a = 0.125$)}
%     \label{fig:figure10}
% \end{figure}
We can further comment on the change in the droplet center distances for the various values of the initial droplet separation. It is observed that the droplet centers come closer initially and this rate of approach of the droplet centers is dependent on the initial separation between them. The droplets come closer then separate and for smaller initial separation, the droplet separation distance starts to increase at an earlier time compared to droplets with larger initial separation. This is due to the thermal wake generated by the leading droplet that distorts the thermal field around the trailing drop leading to a retardation force. For certain degrees of separation, this thermal wake is felt in the very early stages of the droplet migration. The overall effect of the droplets coming closer and then separating is shown to be the same for any initial separation distance. This is summarized in \cref{fig:figure22}.
Similar results were obtained by \citet{kalichetty2019} where they showed that in the absence of any imposed flow, the smaller drop comes near the bigger droplet initially but they separate later. The evolution of the droplet shapes is depicted in \cref{fig:figure831}.
% However, in the presence of an imposed flow, we show similar phenomena in the transverse migration velocities as depicted in \cref{fig:figure22}.
We look at how the droplet migration is influenced by the confinement ratio. The confinement ratio is changed by varying the size of the droplets. The studies in the previous section were conducted with the droplet radius($a$) as $0.125H$ giving a confinement ratio of $0.25$. Here we first increase the confinement ratio to $0.5$ by changing the droplet diameter. The separation distance($d_{0}$) is varied such that the maximum possible separation($d_{0}-2a$) between the points on the two droplets is increased progressively to assess the nature of the droplet-droplet interaction forces with changing confinement ratio($Cr$). All previous geometric($e,p$) and physical($Ca, Ma$) are kept similar to previous simulations. 
\begin{figure}[H]
\centering
    \begin{subfigure}[t]{0.5\textwidth}
    \centering
    \includegraphics[width=0.9\textwidth]{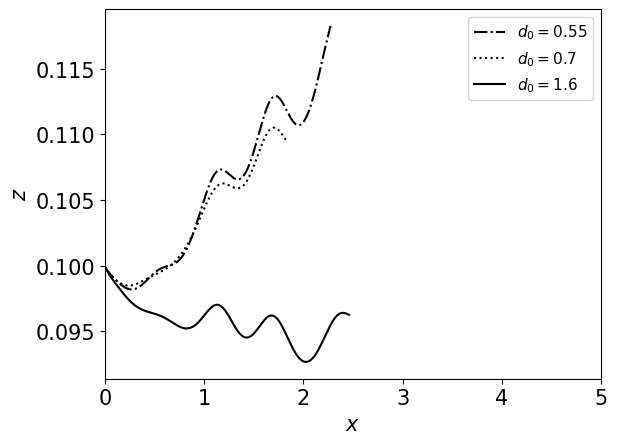}
    \caption{Trailing Droplet}
    \label{fig:figure184}  
    \end{subfigure}%
    \begin{subfigure}[t]{0.5\textwidth}
    \centering
    \includegraphics[width=0.9\textwidth]{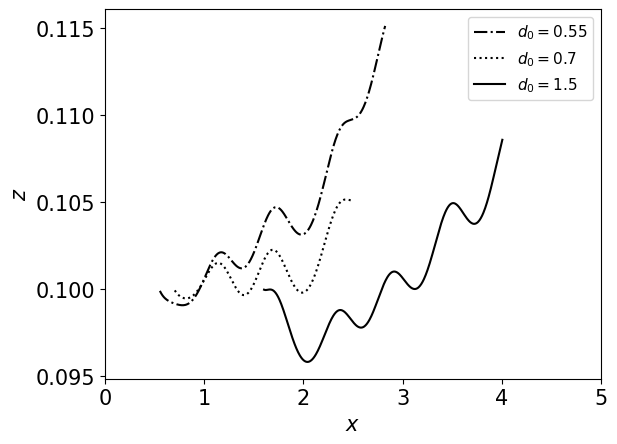}
    \caption{Leading Droplet}
    \label{fig:figure194} 
    \end{subfigure}
    \caption{Droplet Migration trajectories ($a = 0.25$)}
    \label{fig:figure29}
\end{figure}
% \begin{figure}[H]
% \centering
%     \begin{subfigure}[t]{0.5\textwidth}
%     \centering
%     \includegraphics[width=0.9\textwidth]{d_0.5_0.35_0.3_r_0.25_vel.png}
%     \caption{Trailing Droplet}
%     \label{fig:figure185}  
%     \end{subfigure}%
%     \begin{subfigure}[t]{0.5\textwidth}
%     \centering
%     \includegraphics[width=0.9\textwidth]{d_0.5_0.35_0.3_r_0.25_vel_leading.png}
%     \caption{Leading Droplet}
%     \label{fig:figure195} 
%     \end{subfigure}
%     \caption{Droplet Migration velocities ($a = 0.25$)}
%     \label{fig:figure28}
% \end{figure}
Here we see from \cref{fig:figure29} that there is an upward migration of the trailing droplet at smaller separation distances while a downward motion at larger separation distances. The leading droplet keeps migrating toward the wall for all separation distances. 
As the confinement ratio is doubled by increasing the diameters of the droplets and the distance between the wall and the droplets is reduced, we see the changes in the migration trajectories towards the walls. This could be due to the increased droplet interaction forces as a result of the larger droplet size. The effect of these forces is to induce a transverse migration of the trailing droplet towards the wall. For higher separation distances the upward droplet-droplet interaction forces are small and the thermal/droplet-wall forces overpower the droplet-droplet hydrodynamic forces to drive the trailing droplet downward. It can be assumed that in this case of particular conditions, the droplet-wall interaction forces act in the same direction as the thermally induced Marangoni forces in the absence of droplet-droplet interactions. This shows that the only forces driving the trailing droplet upward are the droplet-droplet interaction forces that are dependent on the confinement ratio($Cr$) and the initial droplet separation distance($d_{0}$). The droplet-wall hydrodynamic forces seem to be augmenting the thermal Marangoni forces in driving the trailing droplet in the downward direction. 
% Similar variation in the migration trajectory was described for a single droplet by \citet{das2018}. In their case for the single droplet the migration towards the wall was seen at higher confinement ratios. As the droplet sizes increase the confinement ratio and the wall-induced lift forces increase. 
% This leads to the droplet upward migration since the thermal Marangoni forces are overpowered by the wall forces. 
\par
\begin{figure}[!htb]
    \centering
    \includegraphics[width=0.5\textwidth]{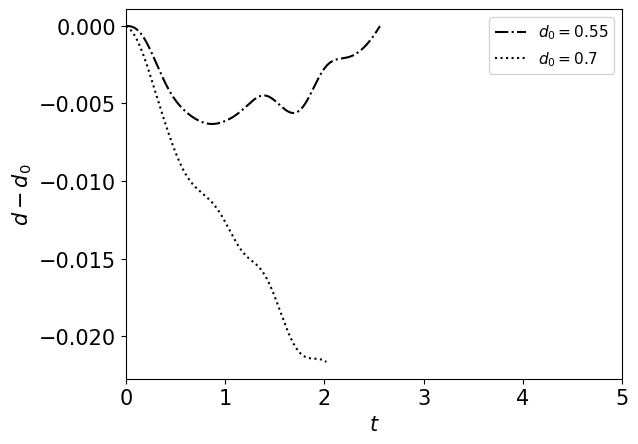}
    \caption{Droplet separation distances with time scale ($a = 0.25$)}
    \label{fig:figure221}
\end{figure}

 Like the previous confinement ratio($Cr$) of $0.25$, we see that the droplets initially come closer and then separate. However, the equilibrium distance is decreased more compared to the case where the droplets are at a lower confinement and thermal wake interactions occur at smaller separation distances, unlike the previous case of the $0.25$ confinement ratio. Thus, it can be seen that the larger size of the droplets giving a higher confinement ratio allows the droplets to come closer further compared to when the confinement ratio is lower. This is due to the reduced effect of the thermal wake that induces the separation of the droplets at higher confinement ratios. At these ratios, the droplet-droplet interaction forces that tend to coalesce the droplets are higher. This leads to droplets coming closer together before separating. 
% The temporal evolution of the velocities is also described in \cref{fig:figure28}. The velocity of both the leading and trailing droplets shows a decrease over time leading to a stable constant amount of separation between the droplets. 
The temporal evolution of the separation of the droplets is depicted in \cref{fig:figure221}. 
Here we increase the droplet radius to $0.35$ such that the confinement ratio becomes $0.7$. Further results are described below.
\begin{figure}[H]
\centering
    \begin{subfigure}[t]{0.5\textwidth}
    \centering
    \includegraphics[width=0.9\textwidth]{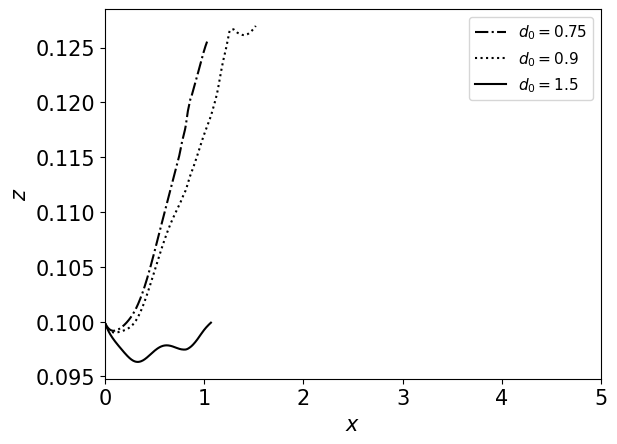}
    \caption{Trailing Droplet}
    \label{fig:figure181}  
    \end{subfigure}%
    \begin{subfigure}[t]{0.5\textwidth}
    \centering
    \includegraphics[width=0.9\textwidth]{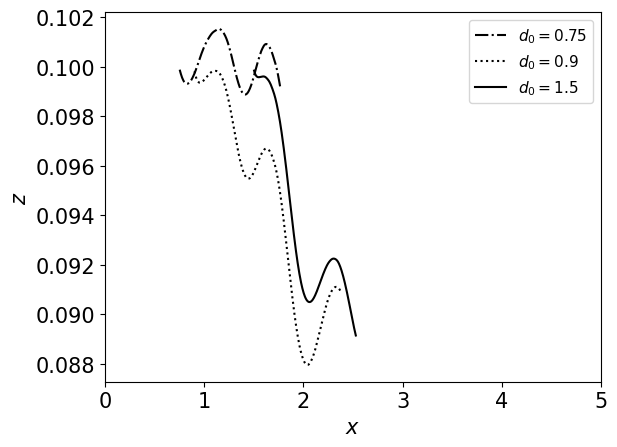}
    \caption{Leading Droplet}
    \label{fig:figure191} 
    \end{subfigure}
    \caption{Droplet Migration trajectories ($a = 0.35$)}
    \label{fig:figure13}
\end{figure}
As seen from \cref{fig:figure13} the trailing droplet migrates towards the wall away from the centreline while the leading droplet migrates towards the centreline. 
The trailing drop migration trajectory is shown until it reaches very close to the insulating walls and is shown to depend on the separation distance between the droplet centers. When the distance of separation is small the migration is faster and vice versa. In the case of the leading droplet, an opposite phenomenon is seen. When the confinement ratio was small, a migration away from the walls and towards the droplet centreline was observed for both the droplets. Here, migration towards the walls has been observed in the case of the leading droplet while in the case of the trailing droplet, a migration towards the centreline is seen. For the trailing droplet, this can be explained in the cases with lower confinement ratios. However, here the motion of the leading droplet can be explained due to the thermal wake created by the trailing droplet as it pushes towards the walls.  
% \begin{figure}[H]
% \centering
%     \begin{subfigure}[t]{0.5\textwidth}
%     \centering
%     \includegraphics[width=0.9\textwidth]{d_0.5_0.35_0.3_r_0.35_vel.png}
%     \caption{Trailing Droplet}
%     \label{fig:figure182}  
%     \end{subfigure}%
%     \begin{subfigure}[t]{0.5\textwidth}
%     \centering
%     \includegraphics[width=0.9\textwidth]{d_0.5_0.35_0.3_r_0.35_leading_vel.png}
%     \caption{Leading Droplet}
%     \label{fig:figure192} 
%     \end{subfigure}
%     \caption{Temporal evolution of the droplet migration velocities ($a = 0.35$)}
%     \label{fig:figure14}
% \end{figure}

% \par The temporal evolution of the migration velocities of both droplets is depicted in \cref{fig:figure14}. It can be seen that for the leading droplet, the migration velocities of both drops reduce with time. However, the leading drop migration velocity reduces with time slower compared to the trailing drop. Moreover, the velocity of the droplets shows a clear variation with the separation distance. In the leading droplet, the migration velocities vary inversely with the separation distance while in the case of the trailing droplet, the effect is reversed. 
\par The separation distance between the droplets is also shown to decrease and then remain constant for a particular case of separation distance similar to what is observed for droplets with a lower radius. For droplets with an initial separation distance of ($d_{0} = 0.9$) and ($d_{0} = 1.5$), the distance reduces until the trailing droplet reaches the wall. The reduction in separation distance and then the increase to a constant value can be explained by the thermal wake created by the leading droplet. However, this is only observed for one particular case($d_{0} = 0.75$) since for other larger values of this parameter the droplets don't come sufficiently close before the trailing droplet touches the wall. Just like the previous cases($0.25$ and $0.5$) and studies by \citet{kalichetty2019}, the thermal wake tends to retard the trailing droplet and accelerate the leading droplet. However, the effects of this wake are felt at a much more reduced separation distance compared to the previous cases, and this leads to a lower equilibrium distance. This possibly could be due to the effect of the confinement ratio in reducing the thermal wakes created between the droplets that leads to the equilibrium separation distance between the droplets reducing. Increased droplet-droplet interactions due to the higher confinement ratios oppose the forces created by the thermal wakes in bringing the two droplets closer. Thus it can be concluded that for certain confinement ratios and Marangoni numbers, the droplet separation distances reduce to the extent to the point of almost droplet coalescence. 
\begin{figure}[!htb]
    \centering
    \includegraphics[width=0.5\textwidth]{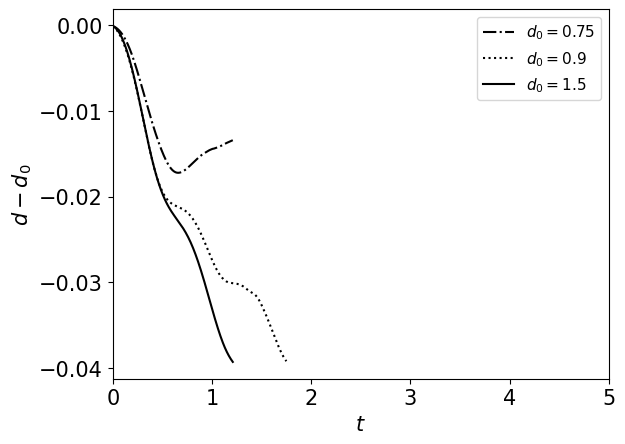}
    \caption{Droplet separation distances with time scale ($a = 0.35$)}
    \label{fig:figure222}
\end{figure}
\begin{figure}[H]
\noindent
\begin{subfigure}{1.0\textwidth}
\centering
\begin{tabular}{|c|c|c|c|}
      \hline
      \addheight{\includegraphics[width=30mm]{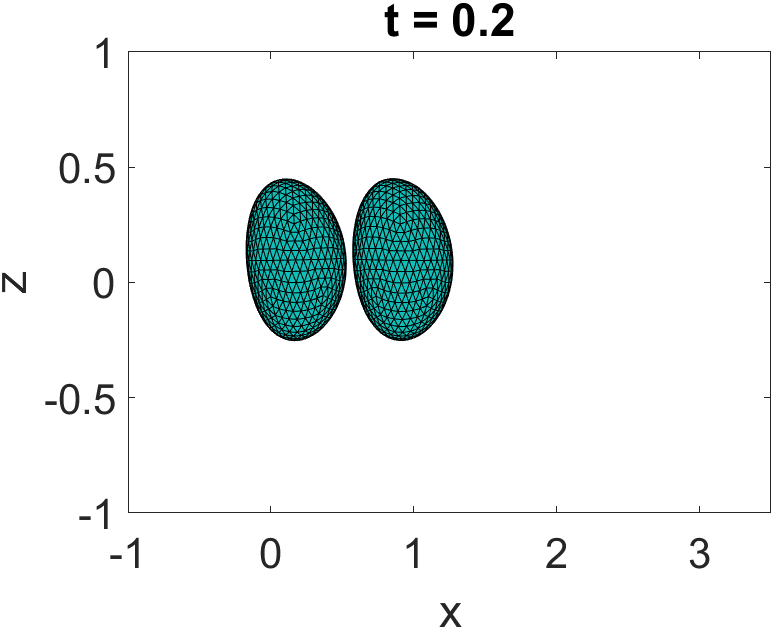}} &
      \addheight{\includegraphics[width=30mm]{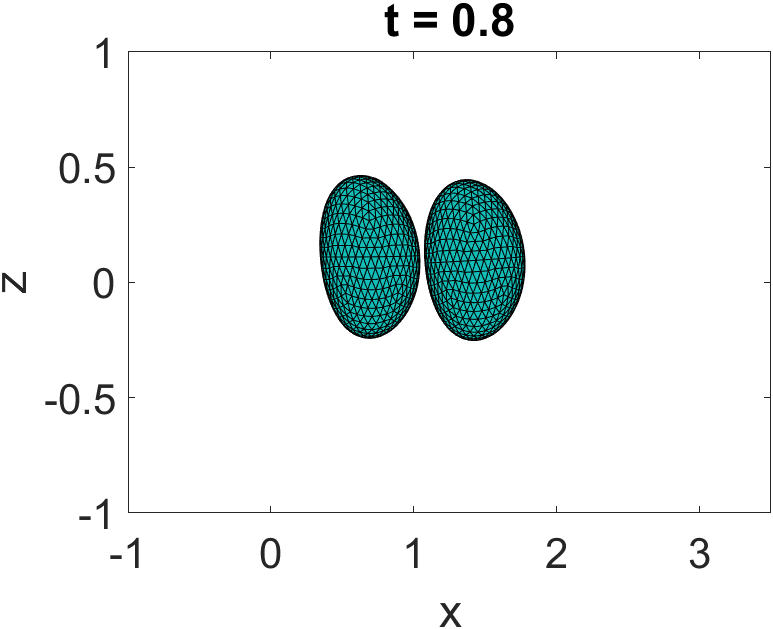}} &
      \addheight{\includegraphics[width=30mm]{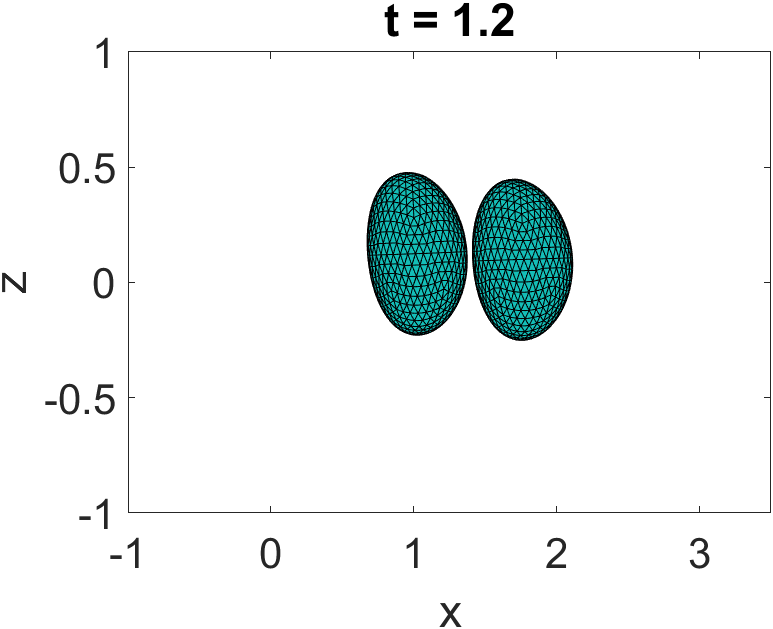}} &
      \addheight{\includegraphics[width=30mm]{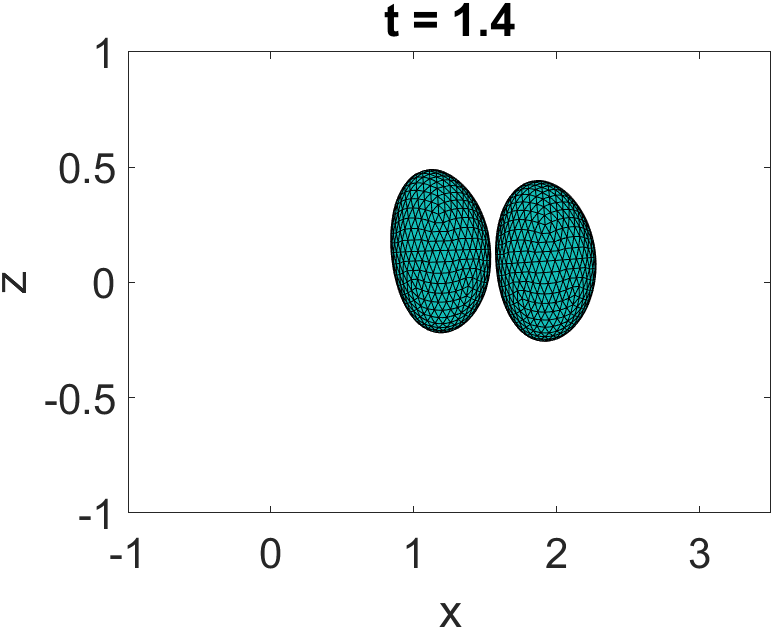}} \\
      \hline     
\end{tabular}
% \caption{evolution of droplet shapes ($a = 0.35,d = 0.75)$}
% \label{fig:figure83}
\end{subfigure}
% \begin{subfigure}{1.0\textwidth}
% \centering
% \begin{tabular}{|c|c|c|c|}
%       \hline
%       \addheight{\includegraphics[width=30mm]{drops_r_0.35_d_0.8,_t_=_0.2.png}} &
%       \addheight{\includegraphics[width=30mm]{drops_r_0.35_d_0.8,_t_=_0.6.png}} &
%       \addheight{\includegraphics[width=30mm]{drops_r_0.35_d_0.8,_t_=_0.8.png}} &
%       \addheight{\includegraphics[width=30mm]{drops_r_0.35_d_0.8,_t_=_1.2.png}} \\
%       \hline     
% \end{tabular}
% % \caption{evolution of droplet shapes ($a = 0.125,d = 0.5)$}
% % \label{fig:figure83}
% \end{subfigure}
\begin{subfigure}{1.0\textwidth}
\centering
\begin{tabular}{|c|c|c|c|}
      \hline
      \addheight{\includegraphics[width=30mm]{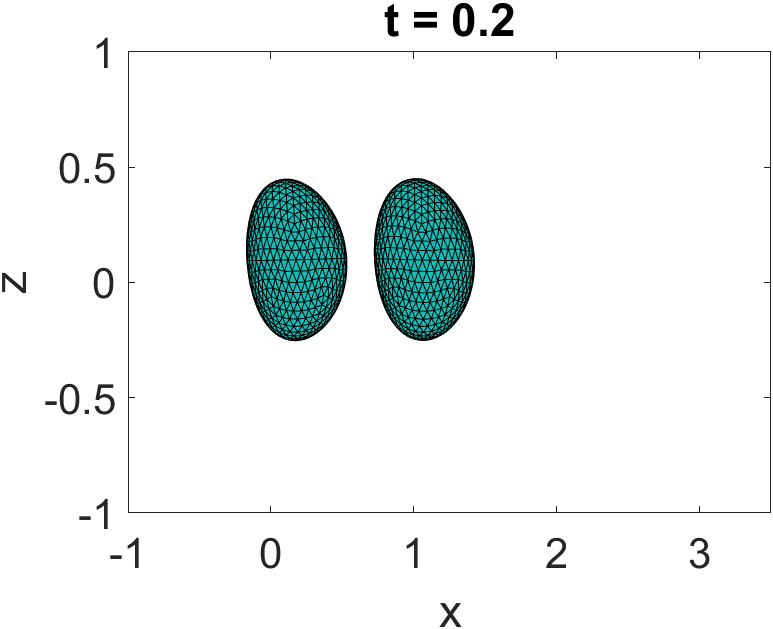}} &
      \addheight{\includegraphics[width=30mm]{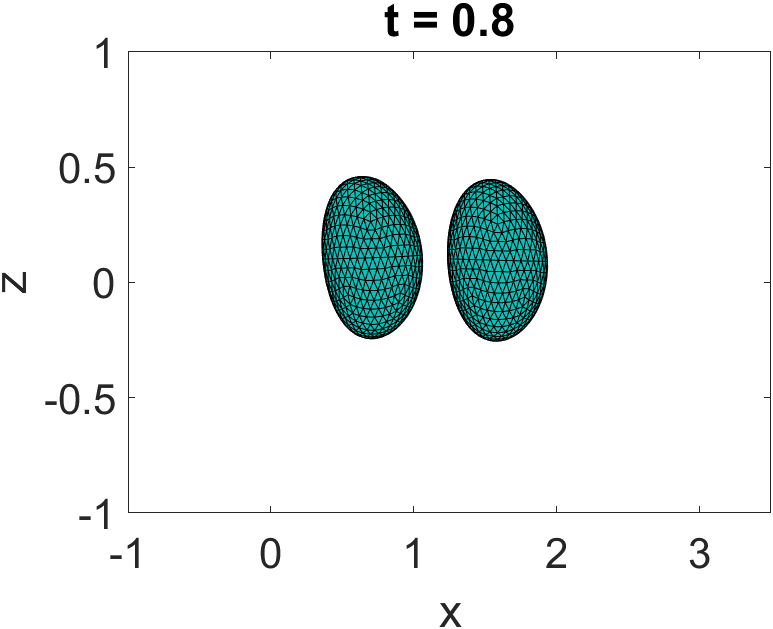}} &
      \addheight{\includegraphics[width=30mm]{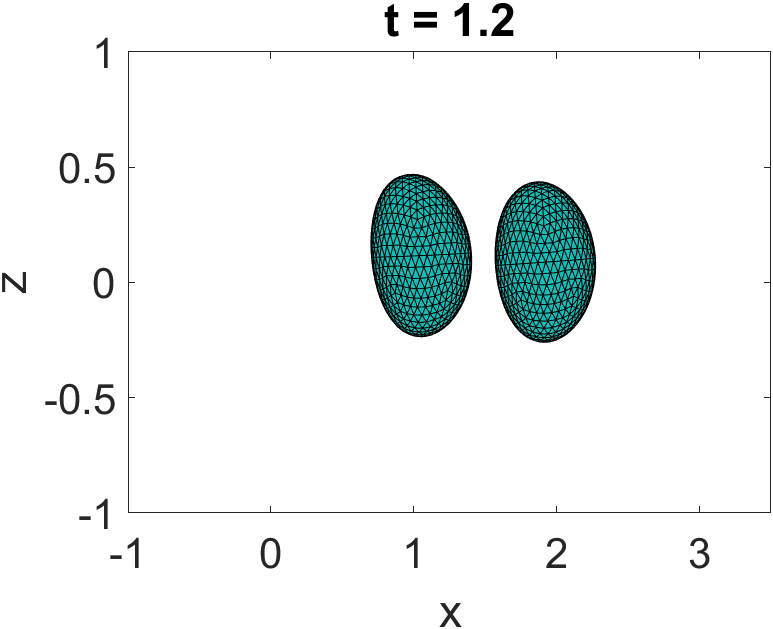}} &
      \addheight{\includegraphics[width=30mm]{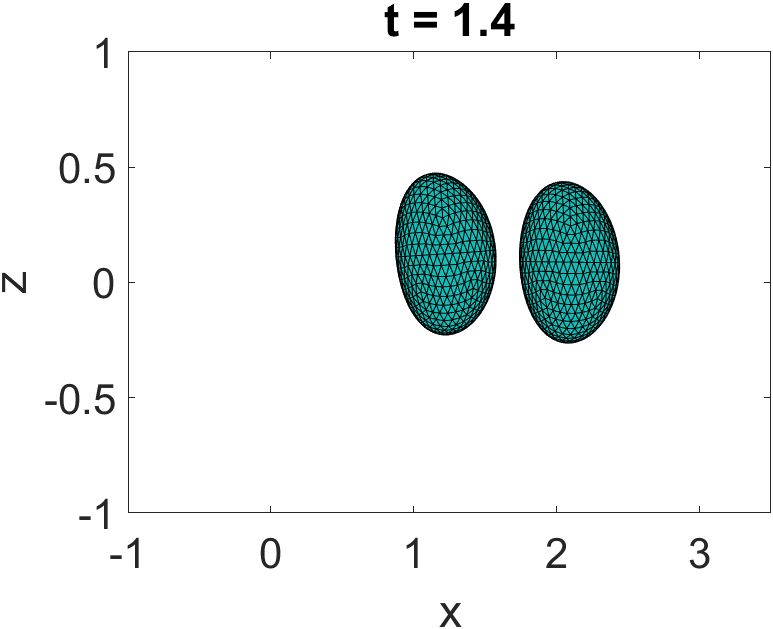}} \\
      \hline     
\end{tabular}
% \caption{evolution of droplet shapes ($a = 0.35,d = 0.9)$}
% \label{fig:figure83}
\end{subfigure}
\caption{Evolution of droplet shapes ($a=0.35$): Top row ($d = 0.75)$ Middle row ($d = 0.8$) Bottom row ($d=0.9$)}
\label{fig:figure83}
\end{figure}
Overall, we see that there are two competing forces acting on the trailing droplet and these are the droplet-droplet interaction forces and the thermal Marangoni forces. The former tends to drive the droplet toward the wall, increases with confinement ratio, and decreases with inter-droplet separation distance. The thermal forces are entirely dependent on the thermal gradients in the flow and are dependent on the Marangoni number. These thermal forces tend to drive the droplet towards the centreline. It is these competing forces that drive the migration of the trailing droplet. The leading droplet migration is dependent on the trailing droplet and the interaction between the hydrodynamic and thermal fields created by the movement of the two droplets.  Droplet deformation is visualized in \cref{fig:figure83}. As can be seen from \cref{fig:figure83} the droplets assume a near bullet-like shape due to the deformation. The droplet-droplet interaction forces are also responsible for bringing the droplets together closer and act opposite to the thermal forces that tend to separate the droplets and prevent coalescence.

\bibliography{main_file}

\end{document}